\documentclass[9pt]{article}
\usepackage{amssymb}
\usepackage{amsfonts}
\usepackage{amsthm}
\usepackage{graphicx}
\usepackage{tabularx}
\usepackage[utf8]{inputenc}
\usepackage{xcolor}
\usepackage[english]{babel}
\usepackage{amsmath}
\usepackage{eucal}
\usepackage{enumerate}
\usepackage{booktabs}
\usepackage{stmaryrd}
\usepackage{amsfonts,epsfig,epstopdf, titling,url, array}
\usepackage{verbatim} 
\usepackage{hyperref}
\usepackage{subcaption}
\usepackage{algorithm}
\usepackage{algpseudocode}
\usepackage{natbib}

\newcommand\bfX{\mathbf{X}}
\newcommand\bfY{\mathbf{Y}}
\newcommand\bff{\mathbf{f}}
\newcommand\bfx{\mathbf{x}}

\newcommand\bfy{\mathbf{y}}
\newcommand\bfz{\mathbf{z}}

\numberwithin{equation}{section}

\newcommand{\beq}{\begin{equation}}
\newcommand{\RN}[1]{%
  \textup{\uppercase\expandafter{\romannumeral#1}}%
}
\newcommand{\eeq}{\end{equation}}

\newcommand{\beqs}{\begin{equation*}}
\newcommand{\eeqs}{\end{equation*}}

\usepackage[
top    = 2.75cm,
bottom = 2.50cm,
left   = 2.49cm,
right  = 2.549cm]{geometry}

\begin{document}
\begin{titlepage}
\LARGE{
\begin{center}
Hybrid SINDy-EnKF in Learning Chikungunya Dynamics from Incomplete, Noisy, or Partially Observed Data
\end{center}
}
\large{
\begin{center} 
Bernard Asamoah Afful\footnote{Corresponding author, e-mail address: bernard.afful@usu.edu}, Changhong Mou\footnote{e-mail address:  a02483619@aggies.usu.edu}, Luis F. Gordillo\footnote{e-mail address:  luis.gordillo@usu.edu} \\[0.3cm]
Department of Mathematics and Statistics, Utah State University, Logan, UT 84322\\
\end{center}
\begin{abstract}
\noindent 
Current mechanistic models for the transmission dynamics of the Chikungunya virus (CHIKV) rely on uncertain parameters or partially observed data. This limitation challenges the use of theoretical models for understanding and forecasting disease spread. Here we present a hybrid, data-driven model framework that combines Sparse Identification of Nonlinear Dynamics (SINDy) with the Ensemble Kalman Filter (EnKF) for sequential data assimilation. Our numerical experiments show that this approach improves prediction accuracy and provides a good reconstruction of unobserved trajectories under partial observability, a common constraint in real-world epidemiological surveillance. SINDy can be applied to epidemic trajectories, recovering the underlying equations in noise-free conditions. However, standalone SINDy is highly sensitive to noise, leading to spurious terms and poor performance.  Hence, we embed the identification procedure within an EnKF framework, which assimilates noisy observations to correct forecast states from the SINDy-derived model and to infer unobserved state variables.
\\[0.3cm]
\noindent
\textbf{Keywords: }Chikungunya virus (CHIKV); Sparse Identification of Nonlinear Dynamics (SINDy); Ensemble Kalman Filter (EnKF); Data assimilation; Infectious disease dynamics; Computational epidemiology.
\end{abstract}
}
\end{titlepage}

\section{Introduction}
Chikungunya virus (CHIKV) is a mosquito-borne arbovirus transmitted primarily by \textit{Aedes aegypti} and \textit{Aedes albopictus}. The spread of CHIKV is considered a significant public health concern in tropical and subtropical regions \cite{morrison, zhao, zhang1}. Compartmental epidemic models, usually formulated via nonlinear ordinary or partial differential equations, have been used to provide insights into the host-vector interactions and intervention strategies for CHIKV. Those models, which are based on epidemiological theory, may allow the derivation of key quantities such as the basic reproduction number or equilibrium states \cite{diekmann, van, yang}. The effectiveness of the models, however, depends on accurate parameter values, which are often difficult to obtain in practice: epidemiological data are frequently incomplete, noisy, or partially observed. This produces uncertainty in model calibration and reduces the predictive power. A useful alternative is data-driven methods, which enable learning of dynamical relationships directly from observed data without requiring a priori mechanistic knowledge. In this paper, we propose a framework for learning Chikungunya transmission dynamics through sparse identification and sequential data assimilation. Specifically, we use Sparse Identification of Nonlinear Dynamics (SINDy) to discover governing equations from limited or noisy data, while employing the Ensemble Kalman Filter (EnKF) to filter noise and infer unobserved state variables.

In recent years, scientists have been using SINDy as a tool for discovering interpretable dynamical systems from time-series data \cite{brunton}. Essentially, if snapshots of a dynamical system and their time derivatives are provided, SINDy performs sparsity-promoting regression (e.g., LASSO or sparse Bayesian inference) on a library of candidate nonlinear functions. Then it recovers a set of governing equations that balance accuracy with parsimony. Due to its interpretability and ease of implementation, SINDy has been widely applied in diverse scientific domains, such as fluid dynamics \cite{alrazen,fukami}, neuroscience \cite{muhammed,selvitella}, chemical kinetics \cite{guo,prabhu}, stochastic modeling \cite{lenfesty,breda}, and epidemiology \cite{babazadeh,jiang}. On the downside, SINDy may perform poorly for noisy data: the algorithm requires numerical estimation of time derivatives, a step where the effects of noise are amplified before propagating into the sparse regression. Therefore, noise may potentially cause true terms to be discarded or spurious terms to be retained \cite{fasel,kaheman}. This is more notorious in low-data, high-noise regimes, which are typical of real-world observations \cite{fasel2}. Also, performance depends on the chosen candidate library and the sparsity threshold, both of which are problem-specific and difficult to tune without ground truth. This contributes to poor robustness across varying data conditions. Recent variants of SINDy have been proposed: Ensemble-SINDy \cite{fasel2} uses bootstrap aggregation to improve robustness to noise and provide uncertainty estimates, while SINDy-PI \cite{kaheman2} reformulates the implicit problem as a convex optimization problem, making the identification of rational and implicit dynamics substantially more noise-tolerant. Despite these remarkable efforts, reliably recovering dynamics from noisy epidemiological observations remains an open problem.

It is known that, for nonlinear systems, the Ensemble Kalman Filter (EnKF) is a computationally efficient method for sequential state estimation \cite{evensen}. By propagating an ensemble of model states and updating them with observations via the Kalman gain, the EnKF optimally balances model errors and observational noise, effectively filtering measurement uncertainty from observed state variables and inferring unobserved state variables using the state covariance matrix estimated from the ensemble spread. This makes it particularly well-suited for epidemiological applications, where data are often sparse, noisy, or partially observed \cite{abbas, belgraoui, lal}.

Hybrid frameworks that couple SINDy with Kalman-type estimation have largely been investigated outside epidemiology, in engineering and the physical sciences. \citet{rosafalco2024ekf} introduced EKF-SINDy, in which a SINDy model serves as the forecast operator of an Extended Kalman Filter, and the two are coupled for joint state--parameter estimation: by augmenting the state with the SINDy coefficients, the filter corrects model error online and stays accurate even outside the library's training regime. Their tests, a shear building driven by recorded seismograms and a partially observed nonlinear resonator, share the noisy, incomplete character of epidemiological surveillance, and a follow-up extended the coupling to bifurcating systems such as the Lotka--Volterra and Selkov models \cite{rosafalco2024online}. Ensemble Kalman methods have likewise been used to learn data-driven closures from indirect observations \cite{zhang2022ensemble}, relying on the same cross-covariance mechanism that carries information from measured to unmeasured components. Comparatively less attention has been given to pairing SINDy with the \emph{ensemble} Kalman filter for partially observed compartmental models in epidemiology, where derivative data for unobserved states are unavailable and cross-compartment couplings can help drive state reconstruction, a setting this study aims to explore.

Therefore, it is desirable to have robust, data-driven methodologies that model disease transmission dynamics from limited, noisy surveillance data while maintaining model interpretability. In this paper, we propose such a design for the spread of CHIKV. Here we formulate a detailed compartmental model for CHIKV transmission that captures host–vector interactions, serving as both a mechanistic foundation and a benchmark for data-driven discovery, followed by SINDy application to reconstruct the governing dynamics with different levels of observational noise, thus providing a systematic assessment of its robustness to noise in an epidemiological context. Finally, we integrate the EnKF to improve state estimation for both observed and unobserved variables, enabling more reliable inference when surveillance data capture only a subset of epidemiological states.

The paper is organized as follows. Section \ref{model} details the CHIKV compartmental model, followed by an overview of the SINDy framework and noise handling in Section \ref{SINDy}. In Section \ref{EnKF}, we describe the Ensemble Kalman Filter for state estimation. We present our numerical results and evaluate the joint SINDy-EnKF framework in Section \ref{results}, concluding with a summary of key findings and future directions in Section \ref{conclusion}.

\section{Learning the Chikungunya Virus Model }
\subsection{The chikungunya virus (CHIKV) model}\label{model}
\begin{figure}[H]
    \centering
    \includegraphics[width=\linewidth]{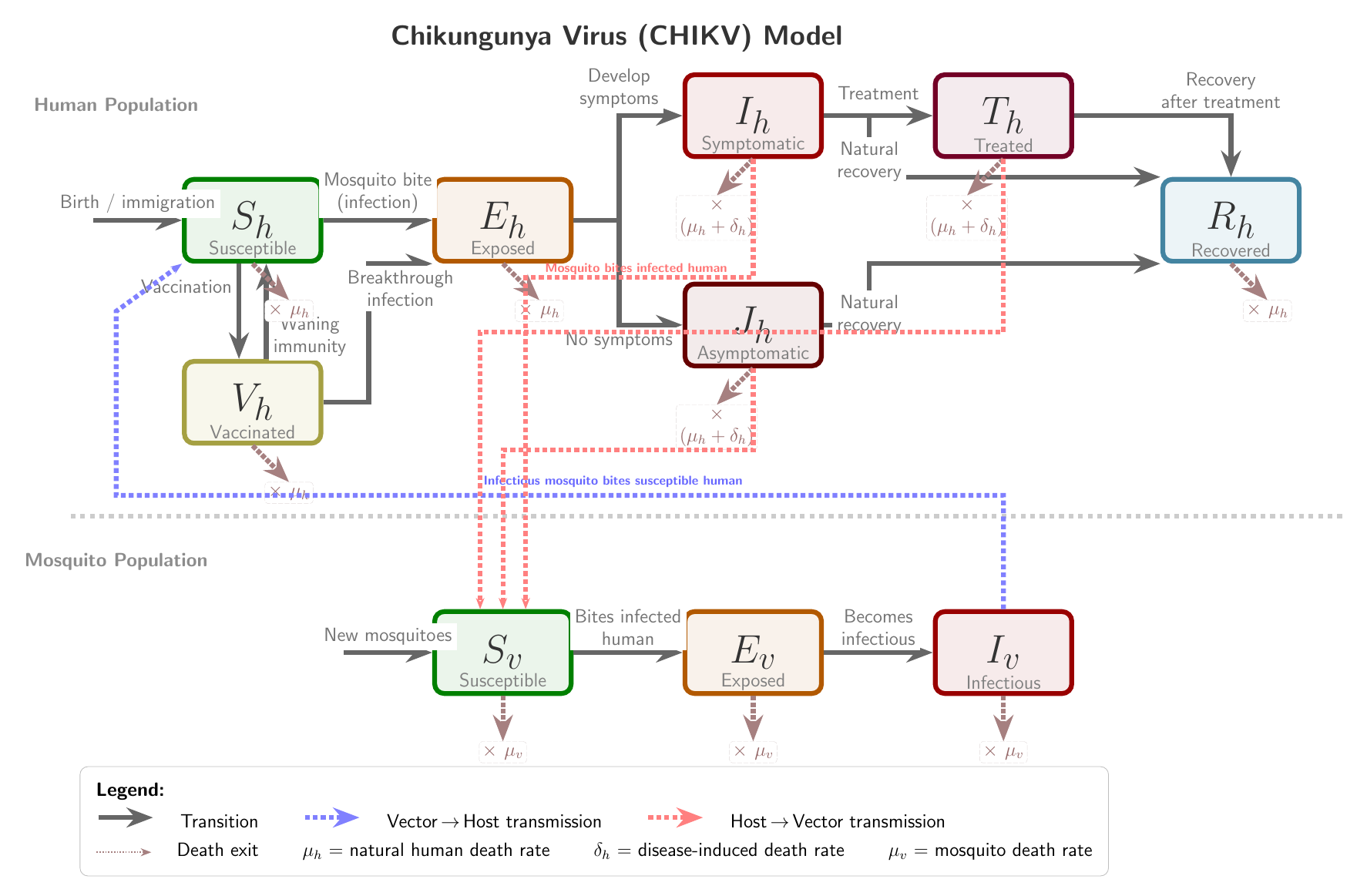}
    \caption{Illustration of the Chikungunya virus (CHIKV) model.}
    \label{fig:chikv}
\end{figure}

The transmission dynamics of the chikungunya virus (CHIKV) are influenced by the mobility of the host and the vector, as well as by the host-vector interaction (human-mosquito) in space and time (\cite{coffey2014chikungunya,guzzetta2020spatial}). In this work, we model these dynamics using a compartmental framework that captures the key epidemiological stages of both populations. All state variables are functions of time $t$, and time dependence is omitted for notational convenience unless otherwise noted. A schematic overview of the model structure, including all compartments and transitions, is shown in Figure \ref{fig:chikv}.
\subsubsection{Host population}
 
The host population $N_h$ is subdivided into seven mutually exclusive compartments: susceptible hosts $S_h$, vaccinated hosts $V_h$, exposed hosts $E_h$, symptomatically infectious hosts $I_h$, asymptomatically infectious hosts $J_h$, treated hosts $T_h$, and recovered hosts $R_h$, so that
\begin{align}
N_h = S_h + V_h + E_h + I_h + J_h + T_h + R_h.    
\end{align}
Susceptible hosts are recruited at a constant rate $\Lambda_h$, and we assume no vertical transmission, meaning all newborns enter the susceptible category. The natural mortality rate of the host population is $\mu_h$, and $\delta_h$ denotes the disease-induced death rate. Susceptible hosts are vaccinated at a rate $\theta$, while vaccine-induced immunity wanes at a rate $\omega$, returning vaccinated individuals to the susceptible class. A susceptible host acquires infection through the bite of an infectious mosquito at a rate governed by the force of infection $\lambda_h = \beta_v I_v / N_v$, where $\beta_v$ is the probability of transmission per bite from an infected mosquito to a host. Vaccinated hosts may also become infected at a reduced rate $\lambda_h(1-\eta)$, where $\eta \in [0,1]$ represents vaccine efficacy, accounting for imperfect vaccine protection. Following an incubation period (typically 2--12 days), exposed hosts transition to either the symptomatically infectious compartment at a rate $\rho\sigma_h$ or the asymptomatically infectious compartment at a rate $(1-\rho)\sigma_h$, where $\rho$ is the probability of developing symptoms. Symptomatically infectious hosts receive treatment at a rate $\tau$. After the viremic period, hosts in the symptomatically infectious, asymptomatically infectious, and treated compartments recover at rates $\gamma_s$, $\gamma_a$, and $\gamma_T$, respectively.
\subsubsection{Vector population}
The vector population $N_v$ is further divided into three mutually exclusive compartments: susceptible vectors $S_v$, exposed vectors $E_v$, and infectious vectors $I_v$, so that
\begin{align}
N_v = S_v + E_v + I_v.
\end{align} 
Susceptible vectors are recruited at a constant rate $\Lambda_v$ and die naturally at a rate $\mu_v$. A susceptible vector acquires the virus through blood feeding on an infectious host at a rate governed by the force of infection
\begin{align}
\lambda_v = \frac{\beta_h(\epsilon_1 I_h + \epsilon_2 J_h + \epsilon_3 T_h)}{N_h},
 \end{align}
where $\beta_h$ is the transmission probability per bite from an infectious host to a mosquito, and $\epsilon_i$ ($i=1,2,3$) are reduction factors that account for differences in transmissibility from symptomatically infectious, asymptomatically infectious, and treated hosts, respectively. After an incubation period, exposed vectors become infectious at a rate $\sigma_v$.

\subsubsection{Governing equations}
Applying the law of mass action and incorporating the assumptions described above \cite{wilson}, we obtain the following system of ordinary differential equations governing the CHIKV transmission dynamics:
\begin{equation}\label{eqn1}
    \begin{aligned}
        \frac{dS_h}{dt}&=\Lambda_h - \frac{\beta_vI_vS_h}{N_v} - \theta S_h + \omega V_h - \mu_h S_h \\
        \frac{dV_h}{dt}&=\theta S_h - \frac{\beta_v(1-\eta)I_vV_h}{N_v} - \omega V_h - \mu_h V_h \\
        \frac{dE_h}{dt}&=\frac{\beta_vI_vS_h}{N_v} + \frac{\beta_v(1-\eta)I_vV_h}{N_v} - (\sigma_h+\mu_h)E_h \\
        \frac{dI_h}{dt}&=\rho\sigma_hE_h - (\tau+\gamma_s+\mu_h+\delta_h)I_h \\
        \frac{dJ_h}{dt}&=(1-\rho)\sigma_hE_h - (\gamma_a+\mu_h+\delta_h)J_h \\
        \frac{dT_h}{dt}&=\tau I_h - (\gamma_T+\mu_h+\delta_h)T_h \\
        \frac{dR_h}{dt}&=\gamma_sI_h + \gamma_aJ_h + \gamma_TT_h - \mu_hR_h \\
        \frac{dS_v}{dt}&=\Lambda_v - \frac{\beta_h(\epsilon_1I_h+\epsilon_2J_h+\epsilon_3T_h)S_v}{N_h} - \mu_v S_v \\
        \frac{dE_v}{dt}&=\frac{\beta_h(\epsilon_1I_h+\epsilon_2J_h+\epsilon_3T_h)S_v}{N_h} - (\sigma_v + \mu_v)E_v \\
        \frac{dI_v}{dt}&=\sigma_v E_v - \mu_v I_v
    \end{aligned}
\end{equation}
with initial conditions:
\begin{align}\label{eqn2}\nonumber
    S_h(0)&=S_{h_0}\geq0,\; V_h(0)=V_{h_0}\geq0,\; E_h(0)=E_{h_0}\geq0,\\
    I_h(0)&=I_{h_0}\geq0,\; J_h(0)=J_{h_0}\geq0,\; T_h(0)=T_{h_0}\geq0,\; R_h(0)=R_{h_0}\geq0,\\\nonumber
    S_v(0)&=S_{v_0}\geq0,\; E_v(0)=E_{v_0}\geq0,\; I_v(0)=I_{v_0}\geq0.
\end{align}

\begin{table}[!ht]
    \caption{Definitions of the parameters and state variables used in the CHIKV model \eqref{eqn1} subject to the initial conditions \eqref{eqn2}.}
    \label{table1}
    \begin{tabular}{lllll@{}}
        \toprule
        State variables& Description\\
        \midrule
        $N_h$, $N_v$ & Total host and mosquito populations \\
        $S_h$, $S_v$ & Total number of susceptible hosts and susceptible vectors \\
        $V_h$ & Total number of hosts who receive vaccinations \\
        $E_h$, $E_v$ & Total number of exposed hosts and vectors\\
        $I_h$, $J_h$ & Total number of symptomatic and asymptomatic infected hosts\\
        $T_h$ & Total number of treated hosts \\
        $R_h$ & Total number of recovered hosts \\
        $I_v$ & Total number of infectious vectors\\
        \midrule
        Parameter& Description& Units\\
        \midrule
        $\Lambda_h$, $\Lambda_v$ & Rates of recruiting susceptible hosts, and female vectors &persons/day, mosquitoes/day\\
        $\mu_h$, $\mu_v$ & Natural death rates of hosts, and vectors& day$^{-1}$\\
        $\delta_h$ & Disease-induced death rate of hosts & day$^{-1}$\\
        $\beta_h$ & Transmission probability per bite (host $\to$ vector)& dimensionless\\
        $\beta_v$ & Transmission probability per bite (vector $\to$ host)& dimensionless\\
        $\epsilon_i,\;i=1,2,3$ & Transmission reduction factor, $\epsilon_i\ll1$& dimensionless\\
        $\omega$ & Waning immunity rate from vaccinated to susceptible & day$^{-1}$\\
        $\theta$ & Vaccination rate & day$^{-1}$\\
        $\eta$ & Vaccine efficacy (0: ineffective, 1: perfect)& dimensionless\\
        $\sigma_h$ & Rate at which exposed hosts become infectious & day$^{-1}$\\
        $\sigma_v$ & Rate at which exposed vectors become infectious & day$^{-1}$\\
        $\rho$ & Probability an exposed host becomes symptomatic& dimensionless\\
        $\gamma_s$ & Recovery rate of symptomatic hosts & day$^{-1}$\\
        $\gamma_a$ & Recovery rate of asymptomatic hosts & day$^{-1}$\\
        $\gamma_T$ & Recovery rate for treated hosts & day$^{-1}$\\
        $\tau$ & Rate of treatment initiation for symptomatic hosts & day$^{-1}$\\
        \bottomrule
    \end{tabular}
\end{table}

\begin{table}[!ht]
    \centering
    \caption{Initial conditions.}
    \label{table3}
    \begin{tabular}{ll|lll@{}}
        \toprule
        State variables& Initial conditions&State variables & Initial conditions\\
        \midrule
        $S_h(0)$ & 1000 & $T_h(0)$ & 0\\
        $V_h(0)$ & 0     & $R_h(0)$ & 0\\
        $E_h(0)$ & 50    & $S_v(0)$ & 2000\\
        $I_h(0)$ & 10    & $E_v(0)$ & 100\\
        $J_h(0)$ & 5     & $I_v(0)$ & 50\\
        \bottomrule
    \end{tabular}
\end{table}

\begin{table}[!ht]
    \centering
    \caption{The parameter values.}
    \label{table2}
    \begin{tabular}{ll|lll@{}}
        \toprule
        Parameter& Value&Parameter& Value\\
        \midrule
        $\Lambda_h$ & 10         & $\epsilon$ & 0.2\\
        $\Lambda_v$ & 10000      & $\rho$     & 0.7\\
        $\mu_h$     & 1/(80*365) & $\gamma_s$ & 1/7\\
        $\mu_v$     & 1/14       & $\gamma_a$ & 1/5\\
        $\delta_h$  & 0.001      & $\gamma_T$ & 1/6\\
        $\beta_h$   & 0.5        & $\tau$     & 0.1\\
        $\beta_v$   & 0.4        & $\sigma_v$  & 0.33 \\
        $\omega$    & 1/(3*365)  &  $\sigma_h$  & 0.2 \\
        $\theta$    & 0.4        & $dt$ (days) & 0.0001 \\
        $\eta$      & 0.7        & & \\
        \bottomrule
    \end{tabular}
\end{table}

\subsection{Sparse identification of nonlinear dynamics (SINDy)}\label{SINDy}
We briefly describe SINDy \cite{brunton}, a data-driven model discovery method for identifying sparse nonlinear models from measurement data. The goal is to recover or discover the governing equations of the CHIKV model \eqref{eqn1} from observed state data. To this end, we rewrite \eqref{eqn1} in compact form as:
\begin{equation}\label{eqn3}
    \frac{d\bfx}{dt}=\bff(\bfx),
\end{equation}
where $\bfx(t) = (S_h, V_h, E_h, I_h, J_h, T_h, R_h, S_v, E_v, I_v)^T \in \mathbb{R}^{K}$ is the state vector and $K = 10$ is the number of compartments.

\subsubsection{Model discovery from clean data}
Suppose we have access to time-series measurements of the state at $m$ time points $t_1, t_2, \ldots, t_m$. We organize these into a state matrix $\bfX \in \mathbb{R}^{m \times K}$ and a corresponding derivative matrix $\dot{\bfX} \in \mathbb{R}^{m \times K}$:
\begin{equation}\label{eqn4}
    \bfX=
    \begin{bmatrix}
        \bfx_1^T(t_1)\\\bfx_2^T(t_2)\\\vdots\\\bfx_K^T(t_m)
    \end{bmatrix},
    \quad
    \dot{\bfX}=
    \begin{bmatrix}
        \dot{\bfx_1}^T(t_1)\\\dot{\bfx_2}^T(t_2)\\\vdots\\\dot{\bfx_K}^T(t_m)
    \end{bmatrix}.
\end{equation}
The key assumption underlying SINDy is that $\bff$ admits a sparse representation in a library of $p$ nonlinear candidate functions:
\begin{equation}\label{eqn5}
    \boldsymbol{\Theta}(\bfx)=
    \begin{bmatrix} 
        1&\theta_1(\bfx)&\theta_2(\bfx)&\cdots&\theta_p(\bfx)
    \end{bmatrix}
\end{equation}
where each column $\theta_j(\bfX)$ evaluates a candidate function (e.g., polynomials or pairwise products) at each time instance. For the CHIKV model \eqref{eqn1}, the relevant candidates include constant, linear, and bilinear terms arising from mass-action interactions between compartments.
The dynamics \eqref{eqn3} can then be approximated as a sparse linear combination of these candidates :
\begin{equation}\label{eqn6}
    \frac{d\bfx(t)}{dt}\approx\boldsymbol{\Theta}(\bfx)\boldsymbol{\Xi},
\end{equation}
where $\boldsymbol{\Xi}$ is a sparse coefficient matrix whose nonzero entries indicate the active terms in each governing equation. Solving \eqref{eqn6} as an unconstrained least-squares problem typically yields spurious nonzero entries. To enforce sparsity, we instead solve the regularized problem:
\begin{equation}\label{eqn7}
    \boldsymbol{\Xi}=\arg\min_{\boldsymbol{\Xi}}\|\dot{\bfX}_k- \boldsymbol{\Theta}(\bfx)\boldsymbol{\Xi}\|_2^2 + \lambda\|\boldsymbol{\Xi}\|_0,
\end{equation}
where $\lambda > 0$ is a regularization parameter promoting sparsity. In practice, \eqref{eqn7} is solved using the Sequentially Thresholded Least Squares (STLSQ) algorithm \cite{zhang}: starting from a least-squares solution, coefficients with magnitude below the threshold $\lambda$ are set to zero, and the least-squares problem is re-solved over the remaining terms. This procedure is repeated for a fixed number of iterations $n$.

\subsubsection{Model discovery from noisy data}
In practice, measurements are usually corrupted by noise. To model this, we define noisy observations as:
\begin{equation}\label{eqn8}
    \bfY=H(\bfX)+ \epsilon^o 
\end{equation}
where $H: \mathbb{R}^{K_s} \to \mathbb{R}^{K_o}$ is an observation operator mapping the full state to $K_o \leq K$ observed components, and $\epsilon^o $ is a zero-mean white noise process with covariance matrix $\mathbf{R}^o$.
The resulting noisy state matrix $\bfY \in \mathbb{R}^{m \times K}$ and its derivative $\dot{\bfY}$ take the same tabular form as \eqref{eqn4}:
\begin{equation}\label{eqn9}
    \bfY=
    \begin{bmatrix}
        \bfy^T(t_1)\\\bfy^T(t_2)\\\vdots\\\bfy^T(t_m)
    \end{bmatrix},
    \quad
    \dot{\bfY}=
    \begin{bmatrix}
        \dot{\bfy}^T(t_1)\\\dot{\bfy}^T(t_2)\\\vdots\\\dot{\bfy}^T(t_m)
    \end{bmatrix}.
\end{equation}
Analogous to \eqref{eqn5}, we construct a candidate library from the noisy data:
\begin{equation}\label{eqn10}
    \boldsymbol{\Phi}(\bfy)=
    \begin{bmatrix} 
        1&\phi_1(\bfy)&\phi_2(\bfy)&\cdots&\phi_p(\bfy)
    \end{bmatrix}
\end{equation}
and approximate the noisy dynamics as:
\begin{equation}\label{eqn11}
    \frac{d\bfy(t)}{dt}\approx\boldsymbol{\Phi}(\bfy)\boldsymbol{\Pi},
\end{equation}
where $\boldsymbol{\Pi}$ is a sparse coefficient matrix. The corresponding regularized regression problem is:
\begin{equation}\label{eqn12}
    \boldsymbol{\Pi}=\arg\min_{\boldsymbol{\Pi}}\|\dot{\bfY}- \boldsymbol{\Phi}(\bfy)\boldsymbol{\Pi}\|_2^2 + \Tilde{\lambda}\|\boldsymbol{\Pi}\|_0,
\end{equation}
where $\tilde{\lambda} > 0$ is the regularization parameter. 

\subsection{Ensemble Kalman Filter}\label{EnKF}
Consider the ODE system \eqref{eqn3} discretized in time, leading to the 
following state-space form:
\begin{align}
    \bfx_{n+1} &= \mathcal{A}_n(\bfx_{n}) + \boldsymbol{\zeta}_n, 
    \label{eq:state_equation} \\
    \bfz_{n+1} &= H_{n+1}(\bfx_{n+1}) + \boldsymbol{\zeta}_n^o, 
    \label{eq:observation_equation}
\end{align}
where $\mathcal{A}_n$ is a (possibly nonlinear) forward operator that advances the state from time $t_n$ to $t_{n+1}$, and $H_{n+1}$ is the observation operator introduced in \eqref{eqn8}. The term $\boldsymbol{\zeta}_n$ is a $K$-dimensional Gaussian white noise with zero mean and covariance $\mathbf{Q}_n$, representing model error, while $\boldsymbol{\zeta}_n^o$ is a $K_o$-dimensional Gaussian white noise with zero mean and covariance $\mathbf{R}^o$, representing observation noise consistent with \eqref{eqn8}.

Denote by $\mathbf{u}_{m+1|m}^k$, $k = 1, \dots, K$, an ensemble of $J$ 
forecast states at time $t_{n+1}$ obtained via Monte Carlo simulation of the 
forward model $\mathcal{A}_n$. The forecast (prior) ensemble mean is:
\begin{equation}\label{eq:forecast_mean}
    \bar{\bfx}_{n+1|n} = \frac{1}{J} \sum_{j=1}^{J} \bfx_{n+1|n}^{(j)},
\end{equation}
where, for notational simplicity, the approximation is written as an equality. 
Similarly, the analysis (posterior) ensemble mean is:
\begin{equation}\label{eq:posterior_mean}
    \bar{\bfx}_{n+1|n+1} = \frac{1}{J} \sum_{j=1}^{J} \bfx_{n+1|n+1}^{(j)}.
\end{equation}

The prior and posterior error covariance matrices, $\mathbf{P}_{n+1|n}$ and 
$\mathbf{P}_{n+1|n+1}$, are estimated from the ensemble as:
\begin{align}
    \mathbf{P}_{n+1|n} &= \frac{1}{J - 1} 
    \mathbf{X}_{n+1|n}\,\mathbf{X}_{n+1|n}^T, 
    \label{eq:prior_covariance} \\
    \mathbf{P}_{n+1|n+1} &= \frac{1}{J - 1} 
    \mathbf{X}_{n+1|n+1}\,\mathbf{X}_{n+1|n+1}^T, 
    \label{eq:posterior_covariance}
\end{align}
where the deviation (anomaly) matrices are defined as:
\begin{align}
    \mathbf{X}_{n+1|n} &= \bigl[ \bfx_{n+1|n}^{(1)} - \bar{\bfx}_{n+1|n},
    \;\dots,\; \bfx_{n+1|n}^{(J)} - \bar{\bfx}_{n+1|n} \bigr], 
    \label{eq:deviation_prior} \\
    \mathbf{X}_{n+1|n+1} &= \bigl[ \bfx_{n+1|n+1}^{(1)} - 
    \bar{\bfx}_{n+1|n+1},\;\dots,\; \bfx_{n+1|n+1}^{(J)} - 
    \bar{\bfx}_{n+1|n+1} \bigr]. 
    \label{eq:deviation_posterior}
\end{align}

When the observation operator $H$ is linear, the Kalman gain takes the form:
\begin{align}
    \mathbf{K}_n &= \mathbf{P}_{n+1|n}\,H^T 
    \bigl( H\,\mathbf{P}_{n+1|n}\,H^T + \mathbf{R}^o \bigr)^{-1} 
    \nonumber \\
    &= \frac{1}{J - 1}\,\mathbf{X}_{n+1|n}\,
    (H\,\mathbf{X}_{n+1|n})^T 
    \left( \frac{1}{J - 1}\,(H\,\mathbf{X}_{n+1|n})\,
    (H\,\mathbf{X}_{n+1|n})^T + \mathbf{R}^o \right)^{-1}. 
    \label{eq:kalman_gain}
\end{align}
where the observation operator $\mathbf{G}$ is assumed to be linear. 
For a nonlinear observation operator, the matrix product 
$H\,\mathbf{X}_{n+1|n}$ is replaced by the ensemble approximation:
\begin{equation}\label{eq:nonlinear_observation}
    \mathbf{Z} = \bigl[ H(\bfx_{n+1|n}^{(1)}) - \bar{\bfz},\;
    H(\bfx_{n+1|n}^{(2)}) - \bar{\bfz},\;\dots,\;
    H(\bfx_{n+1|n}^{(J)}) - \bar{\bfz} \bigr],
\end{equation}
where $\bar{\bfz} = \frac{1}{J}\sum_{j=1}^{J} H(\bfx_{n+1|n}^{(j)})$. This 
avoids computing the Jacobian of $H$, as would be required in the extended 
Kalman filter.
The posterior update for the ensemble is then:
\begin{equation}\label{eq:posterior_update}
    \bfx_{n+1|n+1}^{(j)} = \bfx_{n+1|n}^{(j)} + \mathbf{K}_n 
    \bigl( \bfz_{n+1}^{(j)} - H(\bfx_{n+1|n}^{(j)}) \bigr),
\end{equation}
where the observation is perturbed as 
$\bfz_{n+1}^{(j)} = \bfz_{n+1} + \boldsymbol{\eta}_{n+1}^{(j)}$, with 
$\boldsymbol{\eta}_{n+1}^{(j)} \sim \mathcal{N}(\mathbf{0}, \mathbf{R}^o)$. This stochastic perturbation ensures an asymptotically correct estimate of the analysis error covariance for large ensembles. However, for small ensemble sizes, this perturbation introduces additional sampling errors that affect the covariance estimate.
\begin{figure}[H]
    \centering
    \includegraphics[width=.95\linewidth]{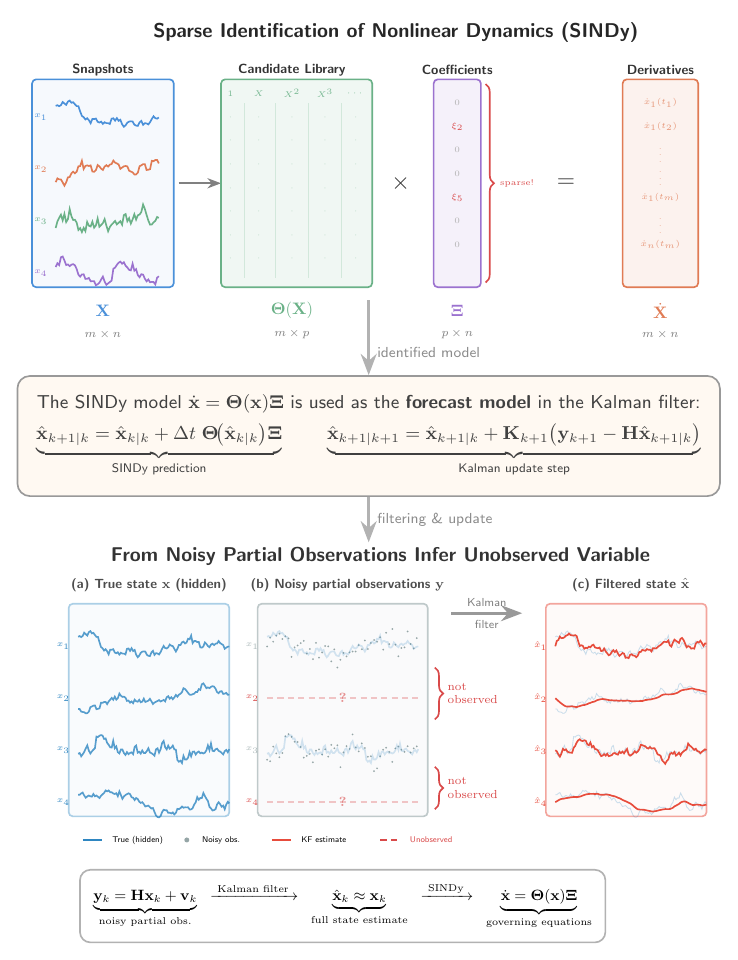}
    \caption{Illustration of learning chikungunya transmission dynamics proposed in this paper: a sparse identification and sequential data assimilation approach.}
    \label{fig:placeholder}
\end{figure}

\section{Numerical Results}\label{results}
In this section, we conduct a series of numerical experiments to assess the performance of the proposed data-driven framework for the CHIKV transmission model~\eqref{eqn1}. Our objectives are threefold: (i) to verify that SINDy can faithfully recover the governing equations under idealized, noise-free conditions; (ii) to characterize the degradation of SINDy-based identification as the level of observational noise increases, thereby exposing the practical limitations of pure sparse regression; and (iii) to show that the EnKF, when coupled with the SINDy-discovered surrogate, infers accurate state estimation and enables the reconstruction of unobserved state variables under realistic partial and noisy observations.
 
The reference (``true'') trajectories are obtained by integrating the CHIKV model~\eqref{eqn1} with the parameters listed in Appendix~\ref{appendixA}, using the fourth-Order Runge-Kutta method (RK4) scheme with a fine time step. Throughout, the accuracy of an estimated trajectory $\widehat{\bfx}(t)$ relative to the true trajectory $\bfx(t)$ is quantified by the component-wise root mean square error (RMSE),
\begin{equation}\label{eqn:rmse_def}
    \mathrm{RMSE}_k \;=\; \sqrt{\frac{1}{m}\sum_{j=1}^{m}\bigl(\widehat{x}_k(t_j)-x_k(t_j)\bigr)^2},\qquad k=1,\ldots,K,
\end{equation}
and by the corresponding relative RMSE,
\begin{equation}\label{eqn:rrmse_def}
    \mathrm{rRMSE} \;=\; \frac{\|\widehat{\bfX}-\bfX\|_F}{\|\bfX\|_F},
\end{equation}
where $\|\cdot\|_F$ denotes the Frobenius norm. For the data-assimilation experiments in Section~\ref{sec:enkf-results}, we additionally report the normalized RMSE (NRMSE) and the temporal correlation between estimated and reference trajectories.
\subsection{Identification of CHIKV Dynamics from Clean Observations}\label{sec:sindy-clean}
We first establish a baseline by applying SINDy to noise-free state trajectories. This idealized setting serves as a sanity check, verifying that the chosen candidate library and sparsity-promoting regression can recover the structural form of~\eqref{eqn1} when the data are uncorrupted.
 
\paragraph{Experimental setup}\label{sec:sindy-clean-setup}
The state matrix $\bfX$ and derivative matrix $\dot{\bfX}$ defined in~\eqref{eqn4} are constructed directly from the reference trajectories. The candidate library $\boldsymbol{\Theta}(\bfx)$ comprises a constant, all linear terms, and all pairwise bilinear products of the ten compartments, consistent with the mass-action structure of compartmental epidemic models. The sparse coefficient matrix $\boldsymbol{\Xi}$ is then obtained by solving~\eqref{eqn7} with the STLSQ algorithm. In the noise-free case, we adopt a very small sparsity threshold $\lambda = 10^{-10}$, since any nonzero coefficient produced by the regression corresponds to a term in the underlying dynamics. The discovered system is simulated by forward-integrating it with the same initial conditions and time horizon used to generate the reference trajectories.
 
\begin{figure}[H]
    \centering
	\includegraphics[width=1\linewidth]{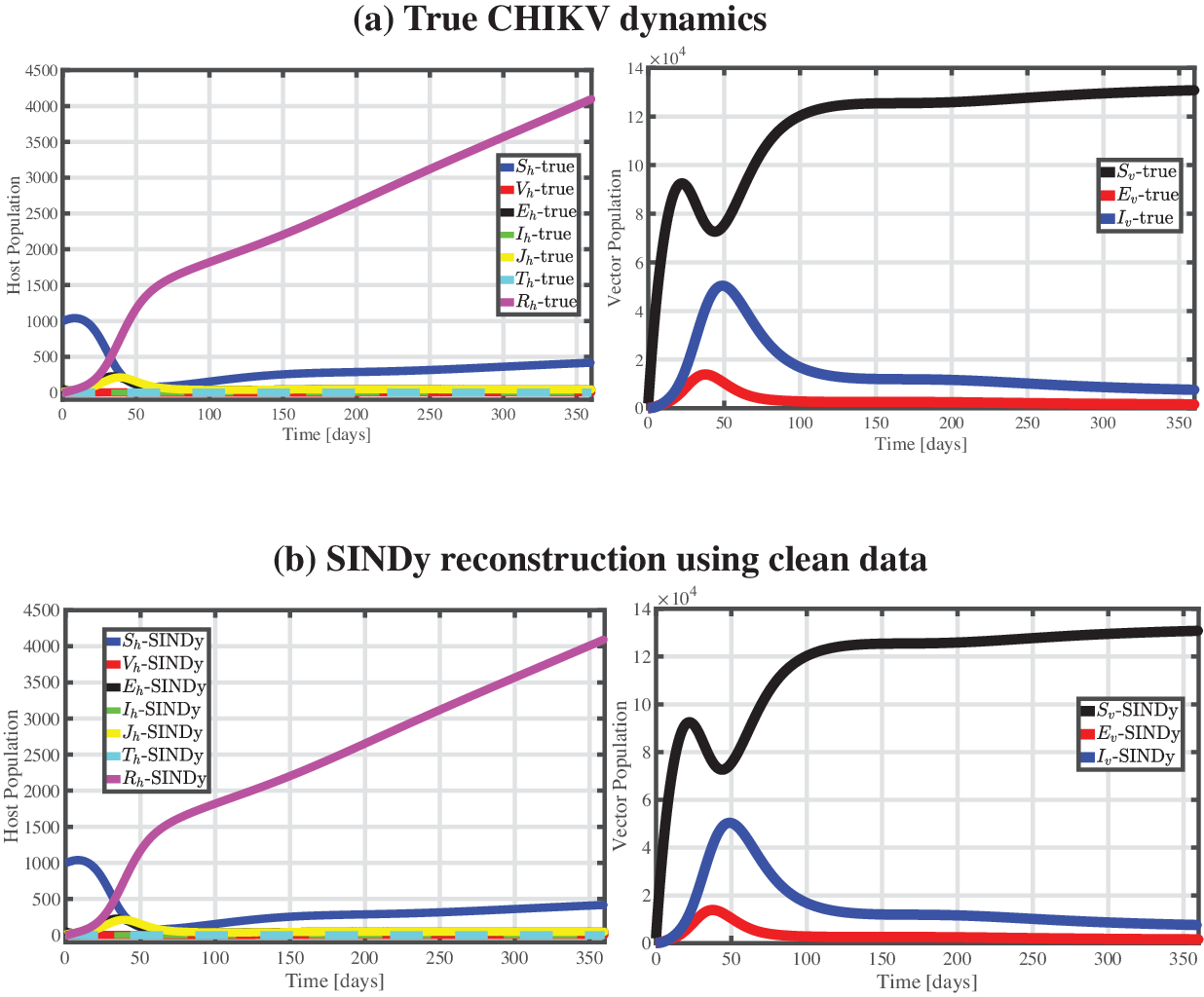}
    \caption{Time evolution of the host and vector compartments obtained by numerically solving the original CHIKV model. The left panel shows the host population dynamics, while the right panel shows the vector population dynamics.}
    \label{fig1}
\end{figure}

\paragraph{Reconstruction accuracy}\label{sec:sindy-clean-accuracy}
Figures~\ref{fig1}(a) and~\ref{fig1}(b) compare the reference and SINDy-reconstructed trajectories for the host and vector populations. The reconstructed dynamics are visually indistinguishable from the truth across all compartments: the initial transient, the epidemic peak, and the long-term equilibrium are accurately reproduced. The quantitative comparisons are reported in Tables~\ref{table4} and~\ref{table5}. Component-wise RMSE values remain below $1$ for every host compartment and below $1$ for the vector compartments relative to populations of order $10^{4}$, and the global relative RMSE is essentially zero ($\mathrm{rRMSE} \approx 10^{-5}$). The aggregate RMSE of $0.2281$ is several orders of magnitude smaller than the characteristic scale of the state variables.
 
\paragraph{Discussion}\label{sec:sindy-clean-discussion}
Two observations are worth emphasizing. First, the equations identified by SINDy (reported in Appendix B) reproduce both the structural form and the parameter values of the original model~\eqref{eqn1} to within numerical precision, indicating that the bilinear candidate library is well adapted to the mass-action transmission mechanism. Second, the parsimony of the discovered model is that only the terms genuinely present in~\eqref{eqn1} have significant coefficients, which further indicates that the STLSQ thresholding successfully suppresses redundant nonlinear monomials despite the large dimensionality of the candidate space.
\begin{table}[ht]
    \centering
    \caption{Component-wise RMSE between the true CHIKV model and the discovered SINDy models. Where $\lambda=0.0000000001$ for clean data, $\lambda=0.00002$ for $5\%$ noise, $\lambda=0.000028$ for $10\%$ noise, $\lambda=0.000001$ for $20\%$ noise, $\lambda=0.0006$ for $25\%$ noise, $\lambda=0.0006$ for $30\%$ noise, $\lambda=0.001$ for $40\%$ noise, and $\lambda=0.005$ for $50\%$ noise.}
    \label{table4}
    \begin{tabular}{|l|c|c|c|c|c|c|c|c|}
        \hline
        \textbf{VAR} & \textbf{Clean data} & \textbf{5\% noise} & \textbf{10\% noise} & \textbf{20\% noise} & \textbf{25\% noise}& \textbf{30\% noise} & \textbf{40\% noise} & \textbf{50\% noise} \\
        \hline
        $S_h$ & 0.0197 & 35.6759 & 96.3218 & 2.4341 & 80.0105 & 56.4741 & 54.4530 & 282.8390 \\
        \hline
        $V_h$ & 0.0001 & 0.0990 & 0.9548 & 0.0246 & 0.8502 & 0.9620 & 0.9561 & 2.0065 \\
        \hline
        $E_h$ & 0.0019 & 5.3416 & 11.2494 & 0.6773 & 18.6239 & 13.5215 & 21.7616 & 50.0348 \\
        \hline
        $I_h$ & 0.0001 & 0.0401 & 0.0023 & 0.0010 & 0.0155 & 0.0159 & 0.0000 & 0.0000 \\
        \hline
        $J_h$ & 0.0020 & 3.1800 & 10.3671 & 0.5330 & 17.1622 & 12.3184 & 20.4717 & 48.3326 \\
        \hline
        $T_h$ & 0.0001 & 0.0032 & 0.0016 & 0.0009 & 0.1170 & 0.0187 & 0.0181 & 0.1087 \\
        \hline
        $R_h$ & 0.0214 & 27.9487 & 282.8673 & 2.3959 & 72.0374 & 78.7419 & 306.1954 & 778.6298 \\
        \hline
        $S_v$ & 0.4893 & 1205.3428 & 2580.0147 & 116.5549 & 12409.5154 & 15878.6285 & 18115.5371 & 19219.8971 \\
        \hline
        $E_v$ & 0.1021 & 185.9490 & 973.8502 & 27.9961 & 983.1011 & 906.3452 & 1706.4929 & 2064.8223 \\
        \hline
        $I_v$ &  0.5194 & 759.2170 & 3686.8817 & 94.1719 & 3344.0028 & 3238.3701 & 6343.3122 & 8083.9721 \\
        \hline
    \end{tabular}
\end{table}

\begin{table}[ht]
    \centering
    \caption{Relative RMSE between the true CHIKV model and the discovered SINDy models. Where $\lambda=0.0000000001$ for clean data, $\lambda=0.00002$ for $5\%$ noise, $\lambda=0.000028$ for $10\%$ noise, $\lambda=0.000001$ for $20\%$ noise, $\lambda=0.0006$ for $25\%$ noise, $\lambda=0.0006$ for $30\%$ noise, $\lambda=0.001$ for $40\%$ noise, and $\lambda=0.005$ for $50\%$ noise.}
    \label{table5}
    \begin{tabular}{|l|c|c|c|c|c|c|c|c|}
        \hline
        & \textbf{Clean data} & \textbf{5\% noise} & \textbf{10\% noise} & \textbf{20\% noise} & \textbf{25\% noise}& \textbf{30\% noise}\\
        \hline
        \textbf{RMSE} & 0.2281 & 454.5252 & 1459.0231 & 48.2178 & 4076.2378 & 5132.7290 \\
        \hline
        \textbf{rRMSE} & 0.0000 & 0.0120 & 0.0385 & 0.0013 & 0.1076 & 0.1355 \\
        \hline\hline
        & \textbf{40\% noise} & \textbf{50\% noise} &  &  &  &\\
        \hline
        \textbf{RMSE} & 6094.4239 & 6631.0606 & &  &  &\\
        \hline
        \textbf{rRMSE} & 0.1609 & 0.1751 &  &  &  &\\
        \hline
    \end{tabular}
\end{table}

\subsection{SINDy Reconstruction from Noisy Data}\label{sec:sindy-noisy}
In practice, epidemiological time series are corrupted by observational noise, sampling fluctuations, and underreporting. To investigate the sensitivity of SINDy to realistic data sets, we systematically add clean trajectories with state-dependent Gaussian noise at different levels.
 
\subsubsection{Noise model and threshold selection}\label{sec:sindy-noisy-setup}
Following~\eqref{eqn8}, noisy observations are generated as
\begin{equation*}
    \bfY = \bfX + \varepsilon\,\sigma(\bfX)\,\mathcal{N}(\mathbf{0},\mathbf{I}),
\end{equation*}
where $\varepsilon \in \{5\%,10\%,20\%,25\%,30\%,40\%,50\%\}$ denotes the relative noise amplitude and $\sigma(\bfX)$ is the empirical standard deviation of each state variable. The sparsity threshold $\tilde{\lambda}$ in~\eqref{eqn12} is tuned for each noise level to balance two competing risks: thresholds that are too small admit noise-driven spurious terms, whereas thresholds that are too large eliminate genuine physical terms. The values used are reported in the captions of Tables~\ref{table4} and~\ref{table5}.

\subsubsection{Small noise regime \texorpdfstring{($\varepsilon \leq 10\%$)}{}}\label{sec:sindy-noisy-low}
For small noise levels, the SINDy reconstruction retains qualitative agreement with the truth (Figs.~\ref{fig3}(a)--\ref{fig3}(b)). The location of the epidemic peak, the relative ordering of compartmental amplitudes, and the long-term decay are all preserved. Nevertheless, quantitative errors begin to accumulate, especially in the vector compartments $S_v$, $E_v$, and $I_v$, which carry the largest absolute populations and are therefore most exposed to additive perturbations. For instance, the RMSE of $S_v$ grows from $0.4893$ (clean) to $1205.34$ at $5\%$ noise and to $2580.01$ at $10\%$ noise (Table~\ref{table4}). Although these errors remain small relative to the equilibrium vector population, which yields a relative RMSE below 4\%, they point to the derivative estimates, rather than the states themselves, as the limiting factor. Numerical differentiation amplifies high-frequency noise components, thereby contaminating the regression targets, even when the state observations appear visually clean.

\subsubsection{Moderate-to-large noise regime \texorpdfstring{($\varepsilon \geq 20\%$)}{}}\label{sec:sindy-noisy-high}
\begin{figure}[!htbp]
    \centering
	\includegraphics[width=1\linewidth]{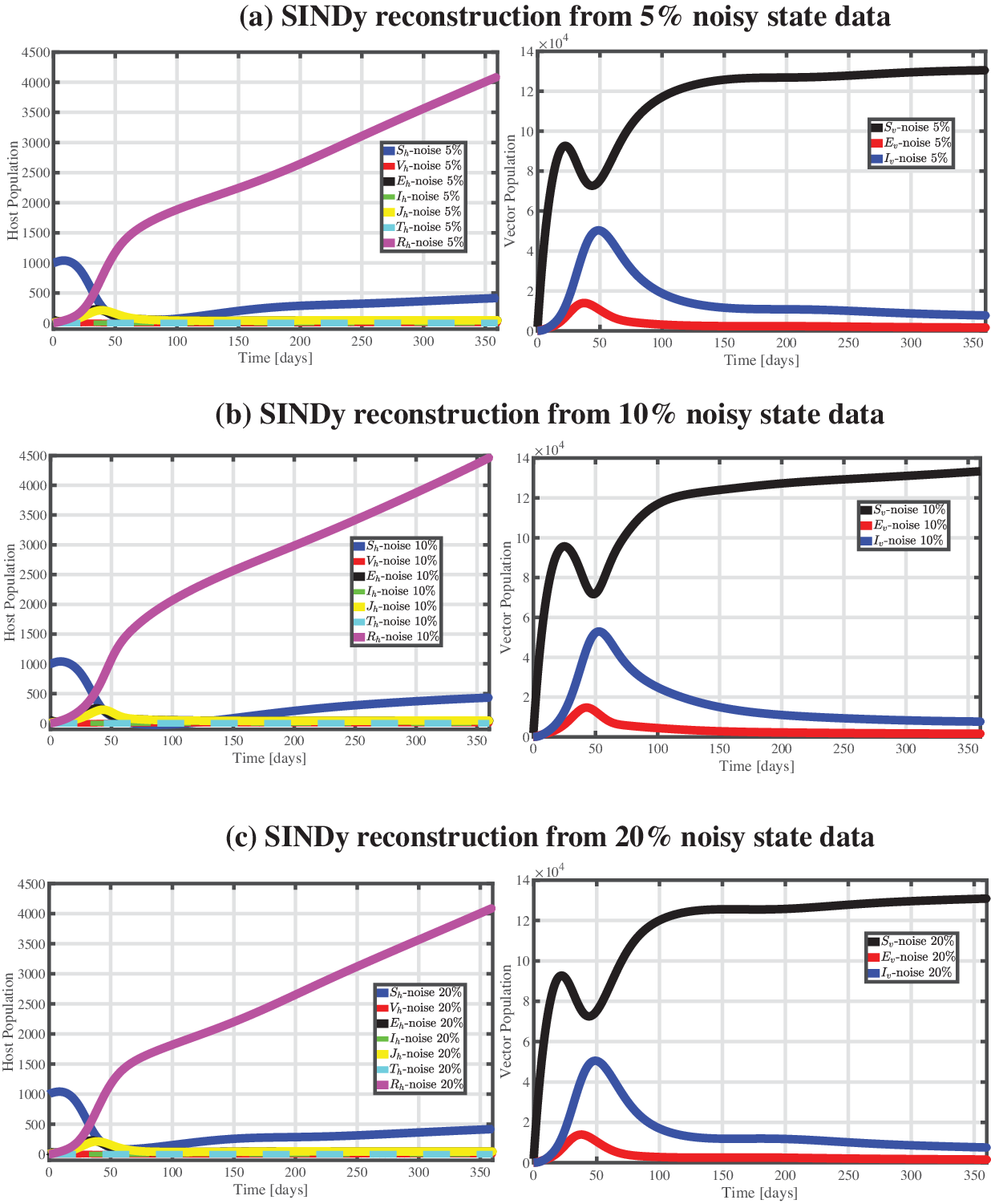}
    \caption{Numerical simulation of the SINDy-identified CHIKV model using noisy state measurements. The left panel shows the host compartments, and the right panel shows the vector compartments.}
    \label{fig3}
\end{figure}

At $20\%$ noise (Fig.~\ref{fig3}(c)), the reconstructed trajectories remain qualitatively consistent with the truth, yet the RMSE is substantially higher than in the clean case. This behavior reflects the coupled effect of noise amplitude, sparsity threshold, and the conditioning of the regression matrix. The aggressive value of $\tilde\lambda$ required to suppress noise-induced terms also eliminates several small-magnitude coefficients, yielding a model that captures the correct qualitative structure but deviates quantitatively in compartments such as $S_v$ and $R_h$.

\begin{figure}[!htbp]
    \centering
	\includegraphics[width=1\linewidth]{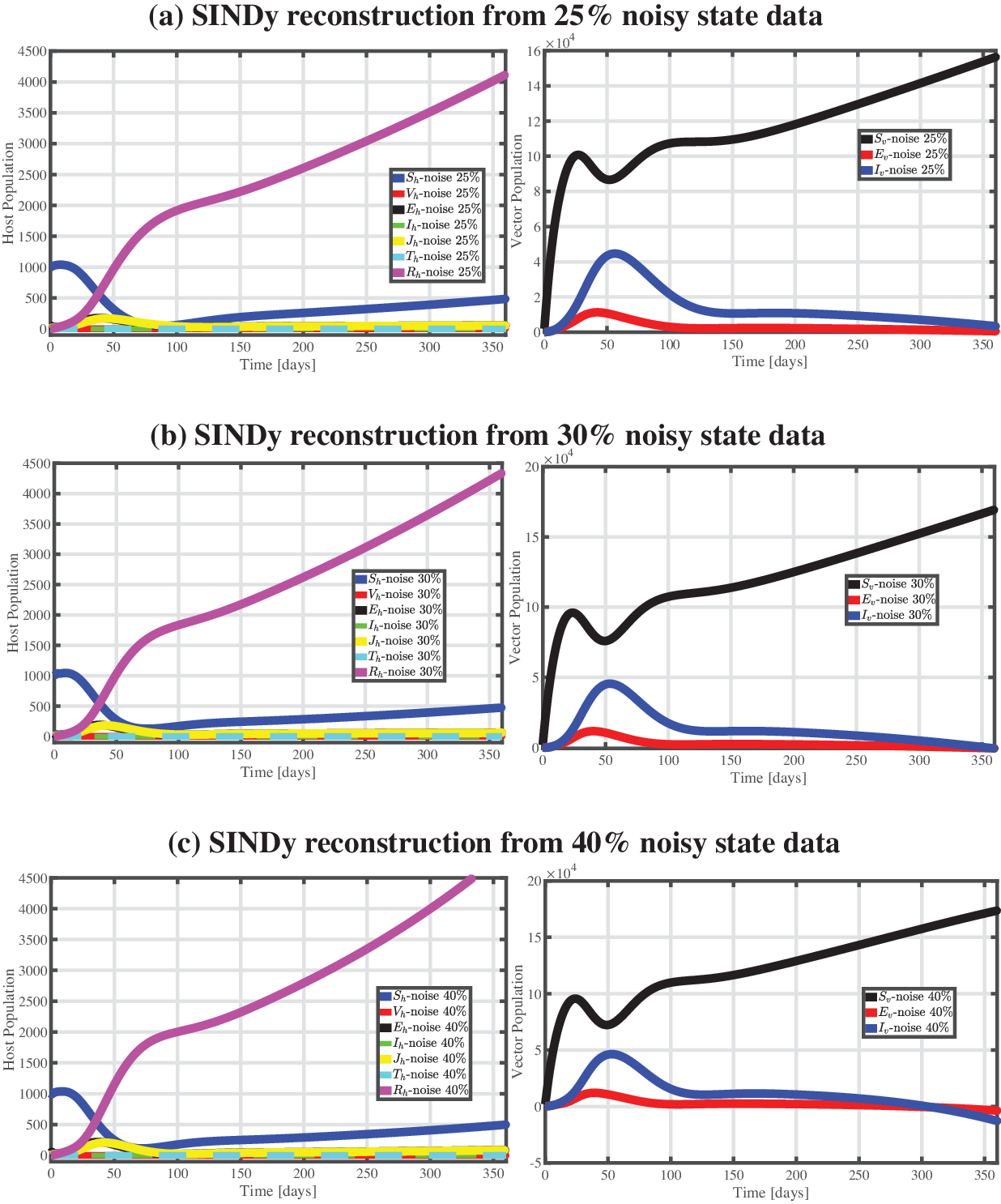}
    \caption{Numerical simulation of the SINDy-identified CHIKV model using noisy state measurements. The left panel shows the host compartments, and the right panel shows the vector compartments.}
    \label{fig6}
\end{figure}

For $\varepsilon \geq 25\%$, the reconstruction degrades consistently (Figs.~\ref{fig6}(a)--\ref{fig6}(c)). Three failure modes appear concurrently: (i) the RMSE increases by three to four orders of magnitude relative to the clean case (Table~\ref{table5}), reaching $\mathrm{rRMSE} \approx 17\%$ at $50\%$ noise; (ii) the identified equations contain nonlinear interactions absent from~\eqref{eqn1} (see Appendix B); and (iii) the resulting dynamics violate biological constraints, with several compartments exhibiting non-physical oscillations or unbounded growth over the integration horizon. The vector compartments are disproportionately affected because their larger absolute scale amplifies state-dependent noise.

The degradation observed here is not specific to the CHIKV model but reflects a general limitation of regression-based identification on time series. SINDy operates on numerically estimated derivatives $\dot{\bfY}$, and finite-difference or smoothing-based differentiation amplifies the noise present in $\bfY$. Once this noise propagates into $\dot{\bfY}$, the least-squares problem~\eqref{eqn12} becomes ill-conditioned with respect to the bilinear library, since several pairwise products are nearly collinear over the observed trajectories. The STLSQ thresholding step then imposes a structural trade-off: a threshold large enough to suppress noise-driven terms also eliminates genuine small-magnitude couplings, whereas a threshold small enough to retain all true couplings admits non-physical terms. The component-wise RMSE values in Table~\ref{table4} quantify this trade-off: compartments with small absolute scale ($I_h$, $T_h$) retain low RMSE at high noise levels because their dynamics are governed by a single dominant flow that persists under any admissible threshold, whereas compartments coupled to multiple flows ($S_v$, $R_h$, $I_v$) accumulate error more rapidly.

Two implications follow. First, a single global sparsity threshold is inadequate for systems whose compartments span several orders of magnitude, motivating the development of compartment-wise or weighted-regression variants of SINDy. Second, and central to the present study, an optimally tuned SINDy model alone is insufficient for trajectory reconstruction in the presence of realistic noise; it must instead be employed as a forecast operator within a data-assimilation framework that sequentially corrects the state using observations. This is the role assigned to the EnKF in the experiments that follow.
\begin{figure}[!htbp]
    \centering
	\includegraphics[width=1\linewidth]{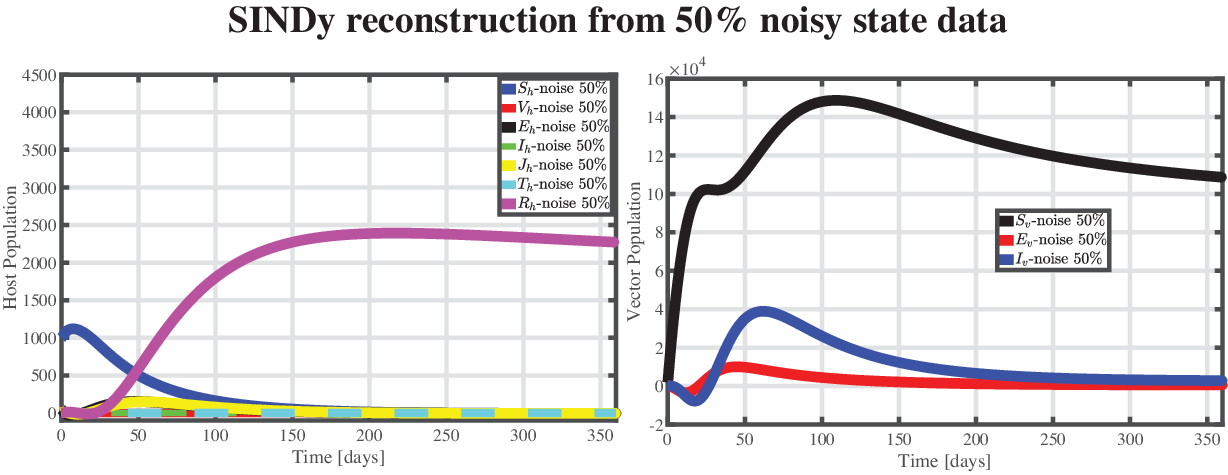}
    \caption{Numerical simulation of the SINDy-identified CHIKV model using noisy state measurements. The left panel shows the host compartments, and the right panel shows the vector compartments.}
    \label{fig9}
\end{figure}
\subsection{State Estimation via Ensemble Kalman Filtering}\label{sec:enkf-results}
We now examine the SINDy--DA approach, in which the SINDy-discovered model serves as the forecast operator $\mathcal{A}_n$ in the state equation~\eqref{eq:state_equation}, and noisy observations are sequentially assimilated to update the state estimate. This formulation addresses two limitations of pure SINDy identification: sensitivity to observational noise, as documented in Section~\ref{sec:sindy-noisy}, and the requirement for full-state observations during training, a condition rarely satisfied in epidemiological surveillance.

\subsubsection{Experimental setup}\label{sec:enkf-setup}
We consider a partially observed setting in which only the host compartments typically reported in surveillance data---vaccinated ($V_h$), infectious ($I_h$), treated ($J_h$), and recovered-under-treatment ($T_h$)---are accessible. The remaining six compartments $\{S_h, E_h, R_h, S_v, E_v, I_v\}$ are treated as unobserved and inferred through assimilation. Observations are perturbed by zero-mean Gaussian noise with covariance $\mathbf{R}^o$ calibrated to a representative noise level. The EnKF is initialized with an ensemble of $N_e$ members drawn from a perturbation of the initial condition, and the forecast is propagated using the SINDy model identified from noisy data. Sensitivity to the observation frequency is assessed by varying the time interval $\mathrm{dt}_{\mathrm{obs}}$ between successive assimilation cycles over the range $[0.1, 10]$.

\subsubsection{Sensitivity to observation frequency}\label{sec:enkf-frequency}
\begin{figure}[!htbp]
    \centering
    \includegraphics[width=1.0\linewidth]{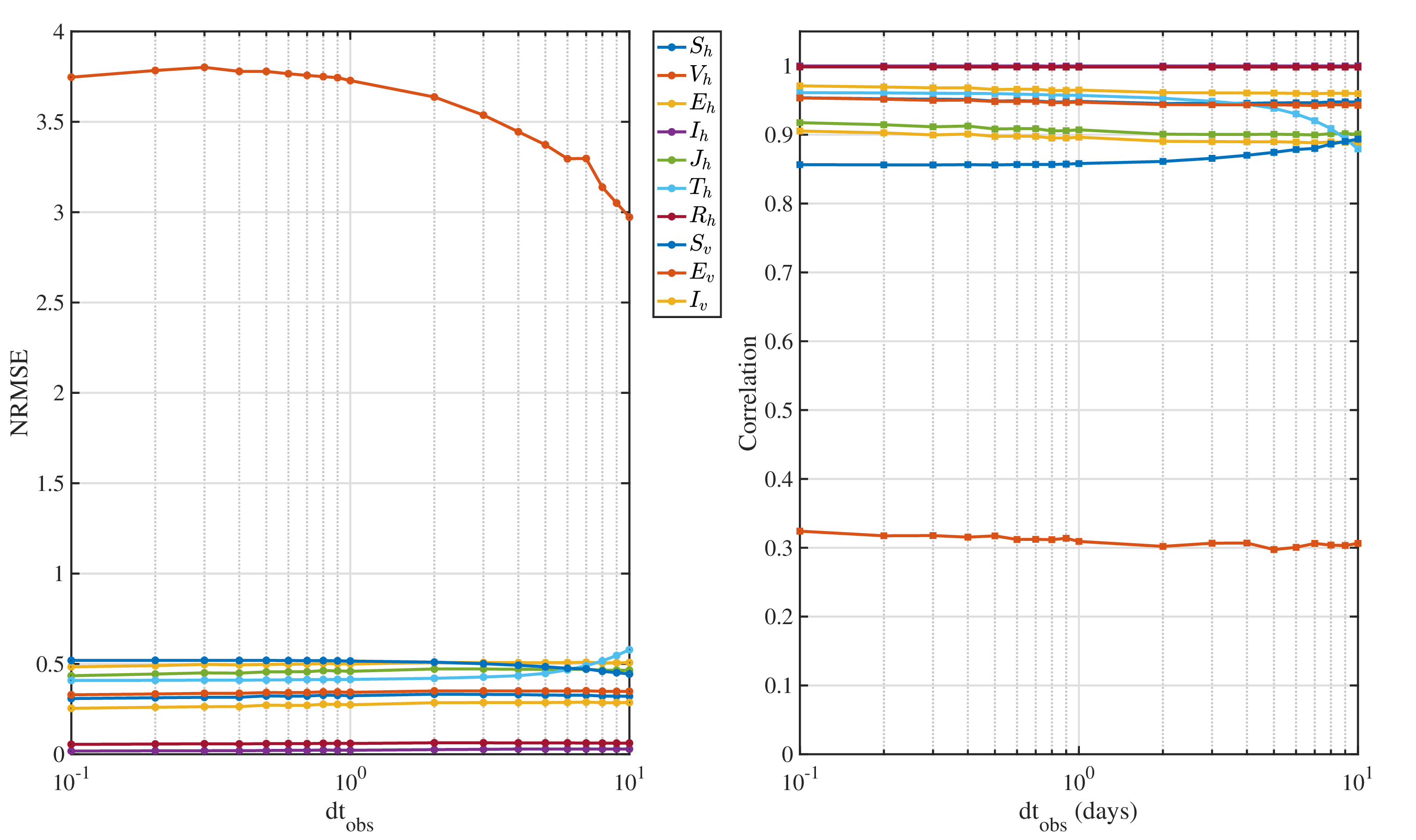}
    \caption{Sensitivity of Ensemble Kalman Filter to the observational time interval $\rm{dt}_{\rm{obs}}$ for the CHIKV model \eqref{eqn1}. The left panel shows the Normalized Root Mean Square Error, while the right panel shows the correlation between the host and vector compartments. Data points represent intervals ranging from 0.1 to 10. Observed variables: $V_h$, $I_h$, $J_h$, and $T_h$.  Unobserved variables: $S_h$, $E_h$, $R_h$, $S_v$, $E_v$, and $I_v$.}
    \label{KF_NRMS}
\end{figure}

Figure~\ref{KF_NRMS} reports the NRMSE and the temporal correlation between the EnKF estimate and the reference trajectory for both observed and unobserved compartments as a function of $\mathrm{dt}_{\mathrm{obs}}$. Three regimes are identified.

In the high-frequency regime ($\mathrm{dt}_{\mathrm{obs}} \leq 1$), the NRMSE remains uniformly low, and the correlation exceeds $0.95$ across nearly all compartments. The assimilation interval is short relative to the characteristic time scales of the host and vector dynamics, so that any drift introduced by the SINDy forecast is corrected before it can amplify. The unobserved compartments are reconstructed with accuracy comparable to that of the observed ones, indicating that the cross-compartment correlations encoded in the ensemble covariance effectively propagate information from the observed subset to the unobserved states.

In the intermediate regime ($1 \leq \mathrm{dt}_{\mathrm{obs}} \leq 5$), the NRMSE increases moderately, and the correlation decreases for unobserved compartments, particularly those weakly coupled to the observed subset. The filter remains stable, but its skill becomes increasingly dependent on the accuracy of the SINDy forecast between assimilation steps.

In the low-frequency regime ($\mathrm{dt}_{\mathrm{obs}} \geq 5$), the estimation error grows substantially. The interval between updates is long enough for the SINDy-driven forecast to accumulate model error, and a single observation is insufficient to reproject the full ten-dimensional state onto the true trajectory. This regime delineates a structural limit of the framework: data assimilation can correct, but does not replace, model accuracy.

\begin{figure}[!htbp]
    \centering
    \includegraphics[width=1.0\linewidth]{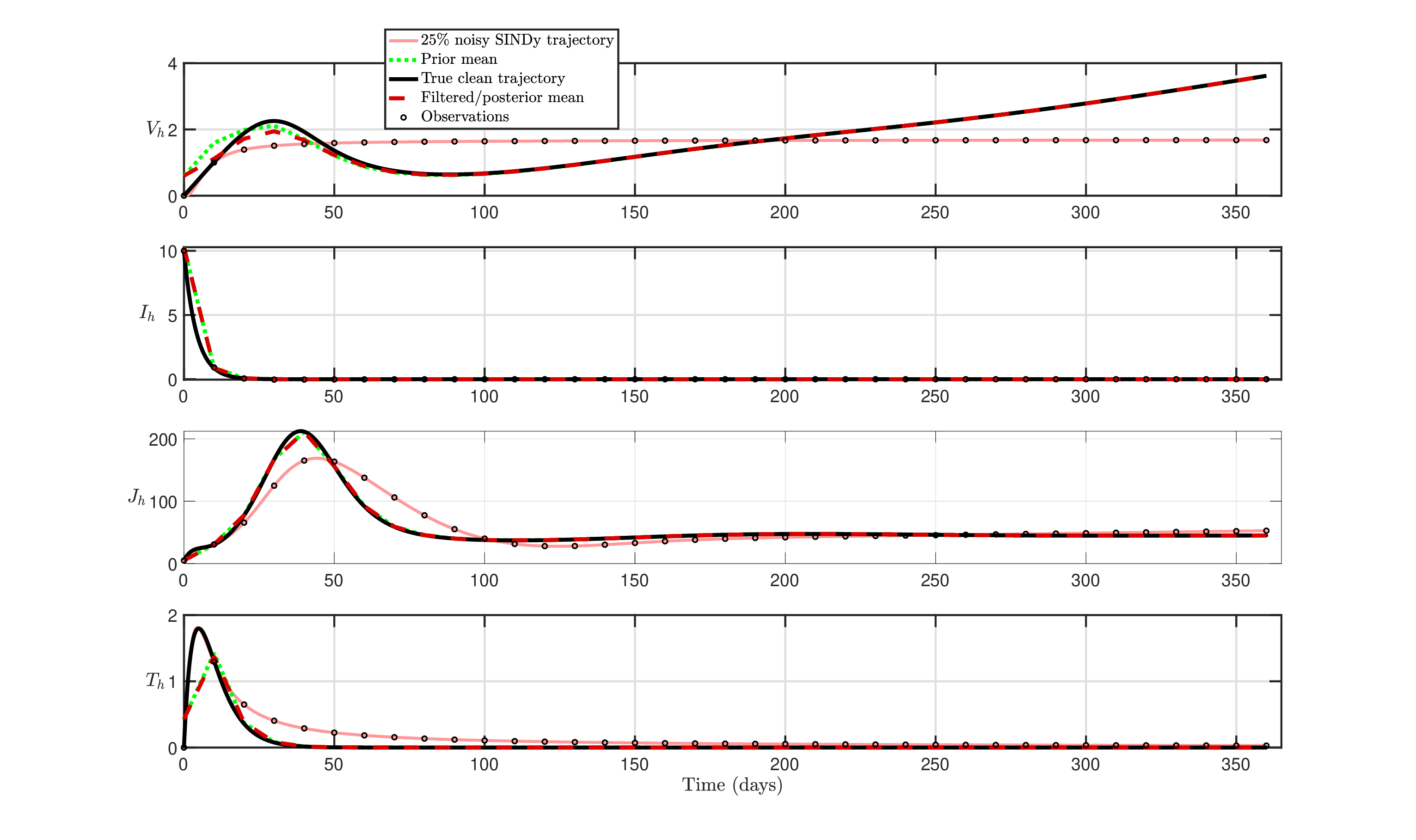}
    \caption{\textbf{EnKF reconstruction of observed host compartments.} Comparison of the $25\%$ noisy SINDy trajectory, forecast prior mean, true clean trajectory, filtered posterior mean, and observations for the observed compartments $V_h$, $I_h$, $J_h$, and $T_h$.}
    \label{observed_KF}
\end{figure}

\begin{figure}[!htbp]
	\centering
	\includegraphics[width=1\linewidth]{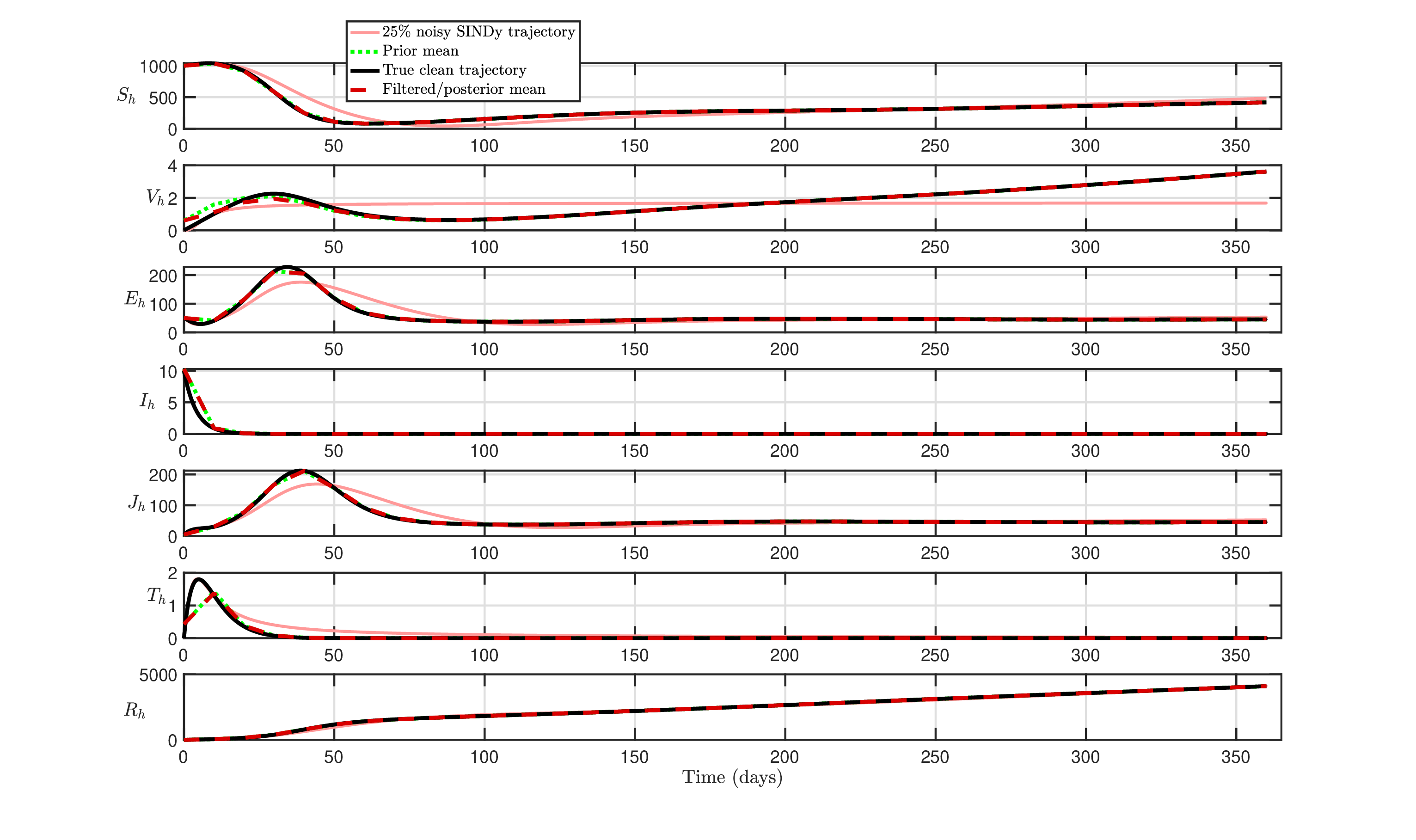}
    \caption{\textbf{EnKF reconstruction of host compartments.} Comparison of the $25\%$ noisy SINDy trajectory, forecast prior mean, true clean trajectory, and filtered posterior mean for all host compartments. The EnKF corrects the noisy SINDy forecast and closely tracks the clean reference dynamics.}
    \label{host_KF}
\end{figure}

\begin{figure}[!htbp]
	\centering
	\includegraphics[width=1\linewidth]{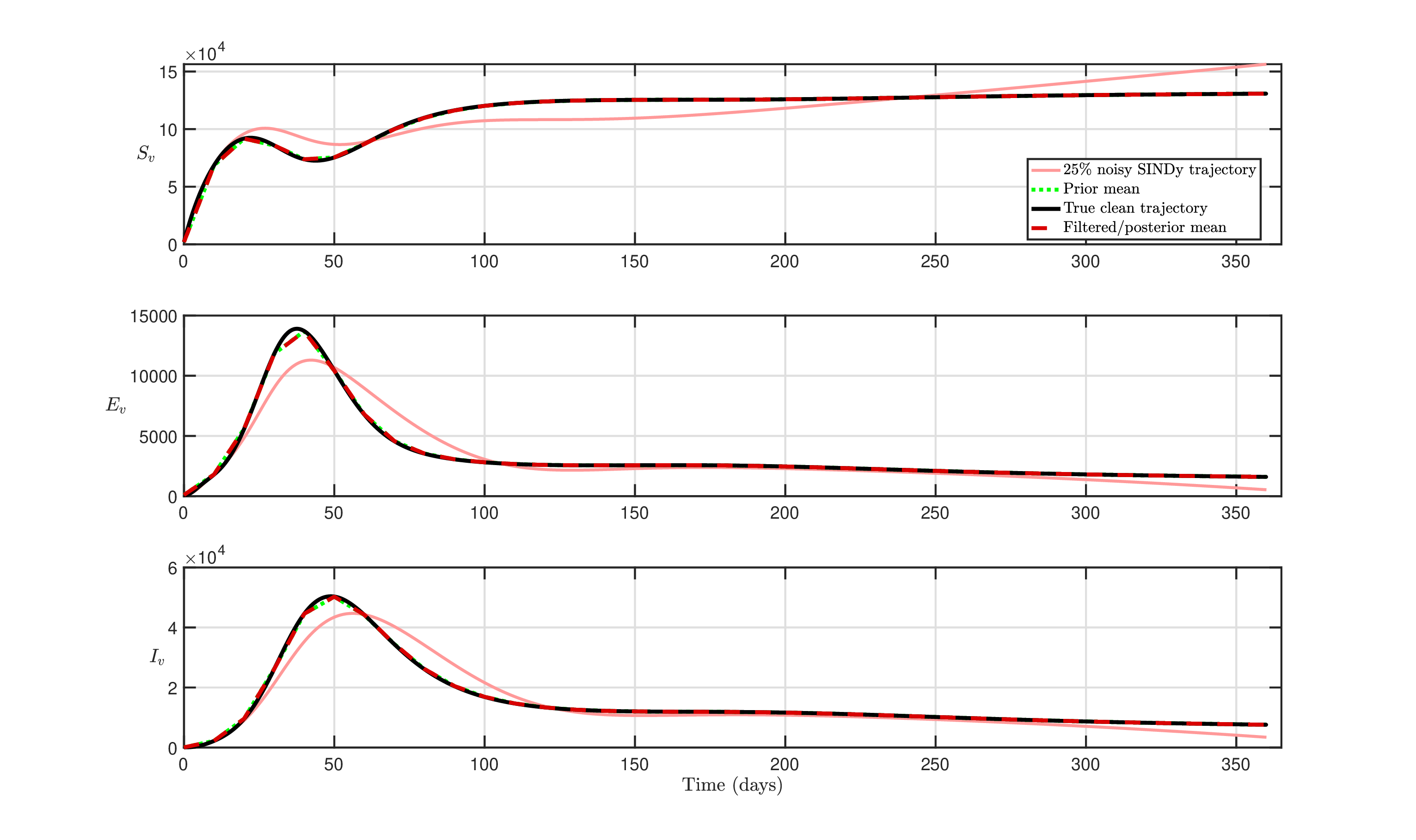}
    \caption{\textbf{EnKF reconstruction of vector compartments.} Comparison of the $25\%$ noisy SINDy trajectory, forecast prior mean, true clean trajectory, and filtered posterior mean for the vector compartments $S_v$, $E_v$, and $I_v$. The filtered estimates remain close to the true clean trajectories despite noise-induced deviations in the SINDy forecast.}
    \label{vectors_KF}
\end{figure}

\begin{table}[ht]
    \centering
    \caption{RMSE comparison between the noisy SINDy trajectory and the SINDy--EnKF posterior estimate.}
    \label{rmse_sindy_enkf}
    \begin{tabular}{lccc}
        \toprule
        State & RMSE SINDy & RMSE SINDy--EnKF & Reduction (\%) \\
        \midrule
        $S_h$ & 79.3176 & 0.3347 & 99.58 \\
        $V_h$ & 0.8690 & 0.1254 & 85.56 \\
        $E_h$ & 18.3900 & 0.1678 & 99.09 \\
        $I_h$ & 0.0122 & 0.0462 & -279.02 \\
        $J_h$ & 16.9518 & 0.1291 & 99.24 \\
        $T_h$ & 0.1155 & 0.0718 & 37.87 \\
        $R_h$ & 71.1182 & 0.9163 & 98.71 \\
        $S_v$ & 12605.0025 & 33.3287 & 99.74 \\
        $E_v$ & 977.5464 & 7.6687 & 99.22 \\
        $I_v$ & 3335.5796 & 27.8186 & 99.17 \\
        \bottomrule
    \end{tabular}
\end{table}

\subsubsection{Recovery of unobserved state variables}\label{sec:enkf-unobserved}
The framework also recovers compartments that are never directly measured---a setting in which SINDy alone is not applicable, since no derivative data are available for those states. The EnKF achieves this by accumulating cross-compartment covariances across the ensemble: information from the infectious host population $I_h$, for example, is propagated through model-induced couplings to update the susceptible vector population $S_v$, even though the latter is not directly observed. Figure~\ref{KF_NRMS} confirms that this indirect inference is reliable for sufficiently frequent observations and degrades smoothly as the observation interval increases.

The robustness of the integrated framework arises from the complementarity of its two components. SINDy provides an interpretable, parsimonious forecast model whose structure encodes the dominant mass-action interactions of the CHIKV system; without such a model, the EnKF would lack a propagator and could not infer unobserved state variables. When SINDy is identified from noisy data, however, the discovered system carries non-negligible model error, while the measurements themselves are subject to observation error. The EnKF reconciles these two sources of uncertainty by producing a statistically optimal state estimate that weights the SINDy forecast and the observations according to their respective error covariances $\mathbf{Q}_n$ and $\mathbf{R}^o$ (see~\eqref{eq:state_equation}--\eqref{eq:observation_equation}). Neither component alone yields reliable trajectory reconstruction under realistic conditions: SINDy degrades under moderate-to-high noise, and the EnKF cannot be deployed without a forecast model. Their combination yields accurate estimates of both observed and unobserved states with quantifiable uncertainty.

Figures~\ref{observed_KF}--\ref{vectors_KF} illustrate the performance of the EnKF when applied to the $25\%$ noisy SINDy-generated trajectories. The noisy SINDy trajectory exhibits visible deviations from the true clean dynamics, especially during the transient phase and near the epidemic peaks. This is most evident in the exposed and infectious vector compartments, where the noisy SINDy trajectory underestimates the peak magnitude and shows noticeable long-time bias.

For the observed host compartments $V_h$, $I_h$, $J_h$, and $T_h$, the filtered posterior mean closely follows the true clean trajectory and substantially improves upon the noisy SINDy trajectory. The observation points coincide with the noisy SINDy trajectory at the assimilation times, but the posterior estimate does not simply interpolate these noisy values. Instead, the EnKF balances information from the forecast model and observations via the Kalman gain, yielding a smoother, dynamically consistent reconstruction.

A particularly important result is the recovery of unobserved compartments. Although $S_h$, $E_h$, $R_h$, $S_v$, $E_v$, and $I_v$ are not directly assimilated, their posterior estimates remain close to the true clean trajectories. This demonstrates that the ensemble covariance successfully transfers information from the observed compartments to the unobserved states. In particular, the vector compartments are reconstructed accurately even though they are not directly observed, indicating that the host--vector coupling encoded in the forecast model provides sufficient dynamical information for indirect state correction.

Table \ref{rmse_sindy_enkf} shows that incorporating the EnKF substantially improves the accuracy of the SINDy forecasts. For most compartments, the RMSE is reduced by more than 98\%, with the largest improvements observed in the vector populations ($S_v$, $E_v$, and $I_v$), where error reductions exceed 99\%. These results demonstrate that sequential data assimilation effectively corrects forecast errors caused by noisy model identification and enables accurate reconstruction of both observed and unobserved states. The only exception is the symptomatic infectious compartment $I_h$, for which the noisy SINDy trajectory already exhibits a very small RMSE. Consequently, minor EnKF corrections lead to a slightly larger error, although the absolute RMSE remains negligible.

These results show that the EnKF corrects the trajectory-level errors introduced by noisy SINDy model discovery. While standalone SINDy is sensitive to noise and may produce biased trajectories, the filtering step suppresses these deviations and recovers biologically plausible dynamics. This supports the use of the hybrid SINDy--EnKF framework for epidemic systems in which only partial and noisy observations are available.

\section{Conclusion}\label{conclusion}
This study developed a data-driven framework to discover and infer the transmission dynamics of the chikungunya virus (CHIKV) by integrating Sparse Identification of Nonlinear Dynamics (SINDy) with the Ensemble Kalman Filter (EnKF). 
The conclusions are two-fold: First, the results show that SINDy accurately recovers the governing equations of the CHIKV model when high-quality, noise-free data are available. In this regime, the identified models reproduce the temporal dynamics of both host and vector populations with high fidelity, confirming the effectiveness of sparse regression in capturing the dominant interactions underlying the system. 
Second, the numerical tests also reveal an inherent limitation of SINDy when applied to noisy epidemiological data. As the level of observational noise increases, the accuracy of the reconstructed dynamics deteriorates due to the sensitivity of derivative estimation and the instability introduced by sparsity thresholding. This leads to spurious nonlinear terms and a loss of interpretability in the identified models, particularly under moderate-to-high noise conditions. These findings highlight that, while SINDy is a powerful tool for model discovery, its direct application to real-world problems requires additional mechanisms to ensure robustness.

The EnKF results further demonstrate that the proposed hybrid framework can mitigate the limitations of standalone SINDy when applied to noisy data. Although SINDy alone becomes sensitive to moderate and large noise levels, coupling it with sequential data assimilation allows noisy forecasts to be corrected using available observations. The posterior estimates closely track the clean reference trajectories and recover both observed and unobserved compartments, including the vector states. Thus, the SINDy--EnKF framework provides a more reliable approach for learning and reconstructing CHIKV dynamics from incomplete and noise-contaminated data.

The RMSE analysis further confirms that the SINDy--EnKF framework reduces state estimation errors by more than 98\% in most epidemiological compartments and by over 99\% in the vector population, demonstrating its effectiveness for reconstructing both observed and unobserved disease states from noisy and incomplete surveillance data.

Several directions extend naturally from the present study. The trade-off between sparsity and accuracy identified in Section~\ref{sec:sindy-noisy} motivates the development of noise-robust extensions of SINDy, including weak-form formulations that bypass explicit derivative estimation, ensemble-SINDy approaches that aggregate sparse models over bootstrapped subsets of the data, and compartment-wise or weighted-regression variants that account for the differing absolute scales of host and vector populations. A second direction is to replace or augment the bilinear candidate library with neural representations of the underlying dynamics. Physics-informed neural networks (PINNs) can encode known conservation and balance constraints of the CHIKV system while learning residual terms directly from data, thereby offering improved robustness to noise compared with library-based identification. Neural ordinary differential equations (Neural ODEs) provide a continuous-time framework in which the right-hand side $\bff(\bfx)$ is parameterized by a neural network and trained jointly with the observed trajectories, making them well-suited to settings where the governing equations admit no parsimonious closed form. To preserve the interpretability that motivated the SINDy formulation, hybrid models---in which a sparse mechanistic core captures the dominant mass-action interactions and a neural network learns residual dynamics or unresolved processes---offer a particularly attractive avenue. Such hybrid architectures can also be embedded as forecast operators within the EnKF, extending the SINDy--EnKF coupling developed here to higher-capacity propagators. On the assimilation side, iterative and localized variants of the EnKF (IEnKF, LETKF) can improve performance under strongly nonlinear dynamics and high state dimension, while joint state--parameter estimation would allow the SINDy coefficients themselves to be updated online as new observations become available, blurring the current separation between offline identification and online assimilation. Finally, the framework can be extended to richer formulations of CHIKV dynamics---including spatial heterogeneity through reaction--diffusion or metapopulation models, stochastic compartmental models that capture demographic and environmental variability, and age- or risk-structured populations---and validated against real surveillance data from past CHIKV outbreaks, with potential applications in outbreak monitoring, real-time forecasting, and public-health decision support.

\subsection*{Availability of data and material}
Real-world data were not used in this study. The code used to generate synthetic data and simulations is available on \href{https://github.com/Bernard0541/Hybrid-SINDy-EnKF-Chikungunya}{GitHub}.
\subsection*{Supporting information}
Appendix A. The CHIKV model \eqref{eqn1} can be equivalently expressed in matrix--vector form. Appendix B. Discovered SINDy equations. Appendix C. Numerical results evaluating the performance of the Sparse Identification of Nonlinear Dynamics (SINDy) framework when trained on partial and noisy data.
\subsection*{Authors' contributions}
B.A.A.: conceptualization, investigation, methodology, software, writing—original draft, writing—review and editing; C.M.: conceptualization, investigation, methodology, supervision, writing — original draft, writing — review and editing; L.F.G.: investigation, supervision, writing — original draft, writing — review and editing. All authors gave final approval for publication and agreed to be held accountable for the work performed there.
\subsection*{Conflicts of Interest}
We declare that no conflicts of interest or financial conflicts exist.
\section*{Funding}
Not applicable.

\appendix
\section*{Appendix A: CHIKV model}\label{appendixA}
The component-wise ODE system (1) presented in Section 2 can be equivalently written in matrix--vector form. Defining the state vector
\begin{align}
\mathbf{x}(t)=
\begin{bmatrix}
S_h & V_h & E_h& I_h& J_h& T_h& R_h& S_v& E_v& I_v
\end{bmatrix}^\top\in\mathbb{R}^{10},
\end{align}
the system takes the compact form
\begin{equation}\label{eqnA1}
\dot{\mathbf{x}}(t)=\mathbf{b}+A(\lambda_h,\lambda_v)\,\mathbf{x}(t),
\end{equation}
where the constant vector is
\begin{align}
\begin{bmatrix}
\Lambda_h & 0& 0& 0& 0& 0& 0& \Lambda_v& 0& 0
\end{bmatrix}^\top,
\end{align}
and the state-dependent system matrix $A(\lambda_h,\lambda_v)\in\mathbb{R}^{10\times10}$ is given by
\begin{equation}
    A(\lambda_h,\lambda_v)=
    \begin{pmatrix}
        -C_1 & \omega & 0 & 0 & 0 & 0 & 0 & 0 & 0 & 0\\
        \theta & -C_2 & 0 & 0 & 0 & 0 & 0 & 0 & 0 & 0\\
        \lambda_h & \lambda_h(1-\eta) & -C_3 & 0 & 0 & 0 & 0 & 0 & 0 & 0\\
        0 & 0 & \rho\sigma_h & -C_4 & 0 & 0 & 0 & 0 & 0 & 0\\
        0 & 0 & (1-\rho)\sigma_h & 0 & -C_5 & 0 & 0 & 0 & 0 & 0\\
        0 & 0 & 0 & \tau & 0 & -C_6 & 0 & 0 & 0 & 0\\
        0 & 0 & 0 & \gamma_s & \gamma_a & \gamma_T & -\mu_h & 0 & 0 & 0\\
        0 & 0 & 0 & 0 & 0 & 0 & 0 & -C_7 & 0 & 0\\
        0 & 0 & 0 & 0 & 0 & 0 & 0 & \lambda_v & -C_8 & 0\\
        0 & 0 & 0 & 0 & 0 & 0 & 0 & 0 & \sigma_v & -\mu_v
    \end{pmatrix}.
\end{equation}
Where $C_1=\lambda_h+\theta+\mu_h$, $C_2=\lambda_h(1-\eta)+\omega+\mu_h$, $C_3=\sigma_h+\mu_h$, $C_4=\tau+\gamma_s+\mu_h+\delta_h$, $C_5=\gamma_a+\mu_h+\delta_h$, $C_6=\gamma_T+\mu_h+\delta_h$, $C_7=\lambda_v+\mu_v$, $C_8=\sigma_v+\mu_v$. Here $\lambda_h = \beta_v I_v / N_v$ and $\lambda_v = \beta_h(\epsilon_1 I_h + \epsilon_2 J_h + \epsilon_3 T_h)/N_h$ are the forces of infection on the host and vector populations, respectively (see Section 2 for parameter definitions). Because $\lambda_h$ and $\lambda_v$ depend on the state~$\mathbf{x}$, the matrix $A$ is itself state-dependent, so \eqref{eqnA1} remains a nonlinear system. The block structure of $A$ reflects the decoupling between host and vector dynamics at the linear level: the upper-left $7\times7$ block governs the host compartments, the lower-right $3\times3$ block governs the vector compartments, and the two populations are coupled exclusively through the force-of-infection terms $\lambda_h$ and $\lambda_v$.

\section*{Appendix B: SINDy identification using partial and noisy data}\label{appendixC}
This appendix presents additional numerical results evaluating the performance of the Sparse Identification of Nonlinear Dynamics (SINDy) framework when trained on incomplete data. Specifically, only the first half of the time-series data is used for model discovery, and the robustness of the method is further assessed in the presence of additive noise (25\% noise level). The resulting models are simulated over the full time horizon and qualitatively compared with expected system behavior. These results highlight the sensitivity of SINDy to both data availability and noise contamination in high-dimensional epidemiological systems.
\begin{figure}[!htbp]
    \centering
	\includegraphics[width=1\linewidth]{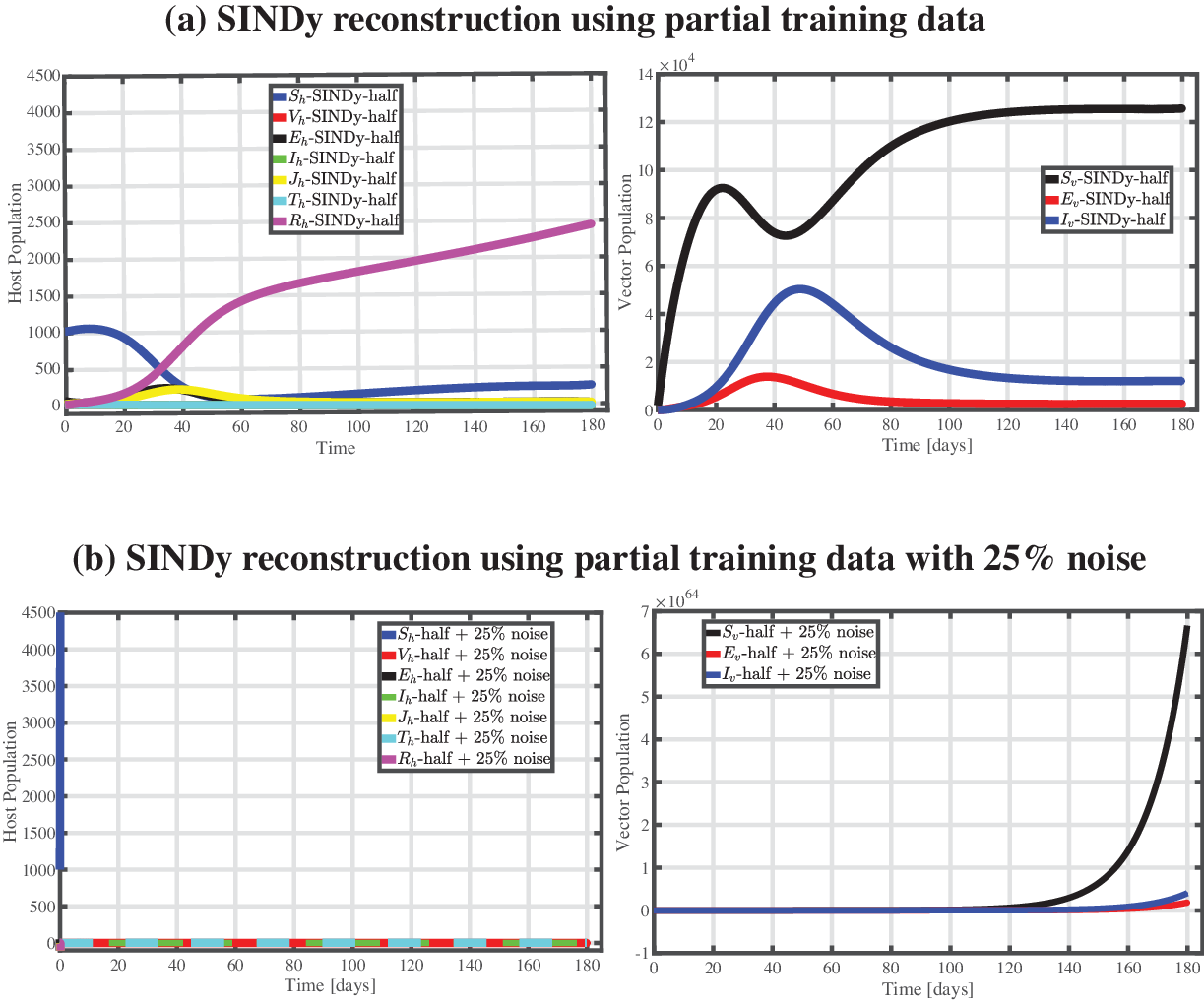}
    \caption{Numerical simulations show the trajectories obtained when SINDy is trained on only the first half of the clean dataset. The left panel shows the reconstructed host dynamics, and the right panel shows the reconstructed vector dynamics.}
    \label{half_host}
\end{figure}

\section*{Appendix C: Discovered SINDy equations}\label{appendixB}
\begin{table}[ht]
    \caption{Discovered SINDy equations with different data.}
    \label{table6}
    \begin{tabularx}{1\linewidth}{|l|X|}
        \hline
        & $\dot{S}_h=-0.0237S_v-0.0005S_h^2+0.0133E_h^2+0.0057J_h^2-0.0003R_h^2+0.0000S_v^2-0.0002I_v^2$ \\
        & $\quad\;\;+0.0014S_hS_v-0.0599V_hE_h+0.0643V_hJ_h+0.0028V_hR_h+0.0001V_hS_v+0.0022V_hE_v$ \\
        & $\quad\;\;+0.0008E_hS_v-0.0663E_hE_v-0.0620I_hJ_h-0.0042I_hR_h-0.0002I_hS_v-0.0009I_hE_v$ \\
        & $\quad\;\;+0.0004I_hI_v+0.0089J_hS_v-0.0462J_hI_v+0.0001T_hE_v$ \\
        & $\dot{V}_h=0.0000$ \\
        & $\dot{E}_h=0.0237S_v+0.0005S_h^2-0.0135E_h^2-0.0055J_h^2+0.0003R_h^2+0.0002I_v^2-0.0015S_hv$ \\
        & $\quad\;\;\;+0.0599V_hE_h-0.0644V_hJ_h-0.0028V_hR_h-0.0001V_hS_v-0.0022V_hE_v-0.0008E_hS_v$ \\
        & $\quad\;\;+0.0664E_hE_v+0.0621I_hJ_h+0.0042I_hR_h+0.0002I_hS_v+0.0009I_hE_v-0.0004I_hI_v$ \\
        & $\quad\;\;-0.0089J_hS_v+0.0461J_hI_v-0.0001T_hE_v$ \\
        & $\dot{I}_h=0.0001S_v$ \\
        & $\dot{J}_h=-0.0001S_v+0.0002E_h^2-0.0003J_h^2+0.0001V_hE_h-0.0001V_hJ_h-0.0001E_hE_v$ \\
        & $\quad\;\;-0.0002I_hJ_h$ \\
        & $\dot{T}_h=0.0000$ \\
        \textbf{Clean data} & $\dot{R}_h=-0.0001S_v+0.0001J_h^2-0.0001V_hE_h+0.0001V_hJ_h-0.0001E_hE_v-0.0001J_hI_v$ \\
        & $\dot{S}_v=-0.1704S_v+0.0092S_h^2+0.0122E_h^2-0.0126*Jh^2+0.0005R_h^2+0.0005E_v^2$ \\
        & $\quad\;\;-0.0006I_v^2-0.0187S_hS_v+0.0740V_hE_h-0.0060V_hJ_h+0.0348V_hR_h+0.0009V_hS_v$ \\
        & $\quad\;\;+0.0007V_hE_v+0.0013V_hI_v+0.0071E_hS_v-0.0058E_hE_v-0.0798I_hJ_h-0.0142I_hR_h$ \\
        & $\quad\;\;-0.0016I_hS_v-0.0047I_hE_v+0.0004I_hI_v+0.0225J_hS_v-0.1255J_hI_v+0.0002T_hR_h$ \\
        & $\quad\;\;+0.0003T_hE_v+0.0001T_hI_v$ \\
        & $\dot{E}_v=-0.0077S_v-0.0013E_h^2+0.0085J_h^2-0.0001R_h^2-0.0003E_v^2+0.0005S_hS_v$ \\
        & $\quad\;\;+0.0162V_hE_h+0.0018V_hJ_h-0.0049V_hR_h+0.0001V_hS_v-0.0005V_hE_v+0.0001V_hI_v$ \\
        & $\quad\;\;+0.0003E_hS_v+0.0070*EhEv+0.0140I_hJ_h+0.0037I_hR_h+0.0002I_hS_v-0.0008I_hE_v$ \\
        & $\quad\;\;+0.0001I_hI_v+0.0020J_hS_v-0.0035J_hI_v-0.0002T_hE_v$ \\
        & $\dot{I}_v=0.0005S_v-0.0001E_h^2-0.0051J_h^2+0.0004E_v^2-0.0002I_v^2+0.0006S_hS_v$ \\
        & $\quad\;\;+0.0035V_hE_h-0.0086V_hJ_h-0.0028V_hR_h+0.0002V_hS_v+0.0002V_hE_v-0.0002V_hI_v$ \\
        & $\quad\;\;-0.0001E_hS_v+0.0063E_hE_v+0.0135I_hJ_h+0.0021I_hR_h-0.0001I_hS_v+0.0007I_hE_v$ \\
        & $\quad\;\;-0.0002I_hI_v+0.0009J_hS_v+0.0063J_hI_v+0.0001T_hE_v$ \\
        \hline\hline
        &  $\dot{S}_h=0.0248S_h+0.0015R_h+0.0010E_v+0.0001I_v-0.0014V_h^2+0.0002E_h^2-0.0278I_h^2$ \\
        & $\quad\;\;-0.0011J_h^2-0.0086T_h^2+0.0739S_hE_h+0.0842S_hJ_h-0.0007S_hR_h-0.0004S_hE_v$ \\
        & $\quad\;\;-0.0003S_hI_v+0.0001V_hE_h+0.4605V_hI_h-0.0003V_hJ_h-0.3950V_hT_h-0.0001V_hR_h$ \\
        & $\quad\;\;+0.2623E_hJ_h-0.0130E_hR_h+0.0004E_hS_v-0.0033E_hE_v-0.0030E_hI_v-0.0003I_hJ_h$ \\
        & $\quad\;\;+0.6727I_hT_h+0.0001I_hR_h+0.0075J_hR_h-0.0001J_hS_v+0.0011J_hE_v-0.0012J_hI_v$ \\
        & $\dot{V}_h=0.0001S_h+0.0002S_hE_h-0.0004S_hJ_h$ \\
        & $\dot{E}_h=-0.0275S_h-0.0001R_h+0.0000S_v-0.0015E_v-0.0001I_v+0.0021V_h^2$ \\
        & $\quad\;\;-0.0004E_h^2+0.0291I_h^2+0.0012J_h^2+0.0098T_h^2-0.0191S_hE_h-0.1507S_hJ_h$ \\
        & $\quad\;\;+0.0008S_hR_h+0.0004S_hE_v+0.0004S_hI_v-0.0005V_hE_h-0.4577V_hI_h+0.0005V_hJ_h$ \\
        & $\quad\;\;+0.4499V_hT_h-0.2774E_hJ_h+0.0135E_hR_h-0.0005E_hS_v+0.0033E_hE_v+0.0030E_hI_v$ \\
        & $\quad\;\;+0.0001I_hJ_h-0.6640I_hT_h-0.0000I_hR_h-0.0089J_hR_h+0.0001J_hS_v-0.0004J_hE_v$ \\
        & $\quad\;\;+0.0010J_hI_v$ \\
        \textbf{5\% noise} & $\dot{I}_h=-0.0002I_h^2+0.0001V_hI_h-0.0006V_hT_h+0.0005E_hJ_h+0.0007I_hT_h$ \\
        & $\dot{J}_h=0.0008S_h-0.0014R_h+0.0000S_v-0.0014E_v+0.0001I_v+0.0004V_h^2+0.0000E_h^2$ \\
        & $\quad\;\;+0.0003I_h^2+0.0978S_hE_h-0.0942S_hJ_h-0.0002V_hE_h+0.0109V_hI_h+0.0003V_hJ_h$ \\
        & $\quad\;\;+0.0211V_hT_h-0.0229E_hJ_h+0.0001E_hR_h+0.0002E_hE_v+0.0001I_hJ_h-0.0750I_hT_h$ \\
        & $\quad\;\;+0.0009J_hR_h-0.0002J_hE_v+0.0001J_hI_v$ \\
        & $\dot{T}_h=0.0001I_h^2-0.0001T_h^2+0.0002V_hI_h+0.0001V_hT_h-0.0001E_hJ_h-0.0004I_hT_h$ \\
        & $\quad\;\;-0.0001J_hR_h$ \\
        & $\dot{R}_h=0.0100S_h+0.0001R_h+0.0023E_v+0.0012V_h^2-0.0007I_h^2-0.0002J_h^2-0.0003T_h^2$ \\
        & $\quad\;\;-0.0017S_hE_h+0.1250S_hJ_h-0.0009S_hR_h-0.0006S_hE_v+0.0001V_hE_h-0.0066V_hI_h$ \\
        & $\quad\;\;-0.0004V_hJ_h-0.1026V_hT_h+0.0684E_hJ_h-0.0005E_hR_h-0.0006E_hE_v+0.0640I_hT_h$ \\
        & $\quad\;\;+0.0018J_hR_h-0.0013J_hE_v$ \\
        & $\dot{S}_v=12.2165S_h+1.5204R_h-0.0125S_v+0.6934E_v+0.0608I_v-0.0031S_h^2+0.8264V_h^2$ \\
        & $\quad\;\;+0.0287E_h^2+0.0783I_h^2+0.0258J_h^2-0.3884T_h^2-0.0008R_h^2-0.0001S_v^2-0.0010E_v^2$ \\
        & $\quad\;\;-11.1410S_hE_h-8.6683S_hJ_h+0.1886S_hR_h-0.0033S_hS_v+0.2817S_hE_v-0.0150S_hI_v$ \\
        & $\quad\;\;+0.0385V_hE_h-3.2008V_hI_h-0.0327V_hJ_h-3.4033V_hT_h+0.0111V_hR_h-0.0002V_hS_v$ \\
    \end{tabularx}
\end{table}
\begin{table}[ht]
    \begin{tabularx}{1\linewidth}{|l|X|}
        & $\quad\;\;+0.0010V_hE_v-0.3487E_hJ_h-0.1068E_hR_h+0.0027E_hS_v+0.0296E_hE_v-0.0148I_hJ_h$ \\
        & $\quad\;\;-5.6657I_hT_h+0.0187I_hR_h-0.0005I_hS_v+0.0026I_hE_v-0.0003I_hI_v-0.1840J_hR_h$ \\
        & $\quad\;\;+0.0061J_hS_v+0.0626J_hE_v+0.0073J_hI_v-0.0004T_hR_h-0.0003T_hE_v-0.0001S_vE_v$ \\
        \textbf{5\% noise} & $\dot{E}_v=0.5005S_h-0.2497R_h-0.0007S_v-0.2900E_v+0.0058I_v-0.0001S_h^2+0.0401V_h^2$ \\
        & $\quad\;\;-0.0162E_h^2-0.0330I_h^2-0.0030J_h^2+0.0737T_h^2-0.0003R_h^2+0.0000E_v^2+4.7099S_hE_h$ \\
        & $\quad\;\;+2.8094S_hJ_h-0.0606S_hR_h-0.0535S_hE_v-0.0029S_hI_v+0.0142V_hE_h+1.4622V_hI_h$ \\
        & $\quad\;\;-0.0129V_hJ_h+1.9717V_hT_h-0.0058V_hR_h+0.0001V_hS_v-0.0010V_hE_v-0.0001V_hI_v$ \\
        & $\quad\;\;+0.1171E_hJ_h+0.0413E_hR_h-0.0009E_hS_v-0.0150E_hE_v+0.0003E_hI_v-0.0111I_hJ_h$ \\
        & $\quad\;\;+3.8198I_hT_h-0.0085I_hR_h+0.0003I_hS_v+0.0006I_hE_v+0.0001I_hI_v+0.0425J_hR_h$ \\
        & $\quad\;\;-0.0024J_hS_v-0.0264J_hE_v-0.0068J_hI_v+0.0001T_hR_h$ \\
        & $\dot{I}_v=0.3300E_v-0.0714I_v$ \\
        \hline\hline 
        & $\dot{S}_h=0.0015S_h+0.0009R_h+0.0023E_v-0.0001I_v-0.0020V_h^2+0.0002E_h^2-0.0170I_h^2$ \\
        & $\quad\;\;-0.0008J_h^2+0.0144T_h^2-0.0657S_hE_h+0.1493S_hJ_h-0.0007S_hR_h-0.0010S_hE_v$ \\
        & $\quad\;\;-0.0002S_hI_v+0.0002V_hE_h+0.6352V_hI_h+0.0002V_hJ_h+0.0252V_hT_h-1.6367E_hI_h$ \\
        & $\quad\;\;-0.0359E_hJ_h+0.0028E_hR_h+0.0001E_hS_v+0.0044E_hE_v-0.0039E_hI_v-0.0006I_hJ_h$ \\
        & $\quad\;\;+0.3265I_hT_h+0.0081J_hR_h-0.0003J_hS_v-0.0036J_hE_v$ \\
        & $\dot{V}_h=0.0001S_h-0.0002S_hE_h$ \\
        & $\dot{E}_h=-0.0045S_h+0.0005R_h-0.0031E_v+0.0001I_v+0.0030V_h^2-0.0003E_h^2+0.0168I_h^2$ \\
        & $\quad\;\;+0.0009J_h^2-0.0109T_h^2+0.0323S_hE_h-0.1455S_hJ_h+0.0011S_hR_h+0.0013S_hE_v$ \\
        & $\quad\;\;+0.0003S_hI_v-0.0004V_hE_h-0.6503V_hI_h+0.0129V_hT_h+1.8145E_hI_h+0.0342E_hJ_h$ \\
        & $\quad\;\;-0.0025E_hR_h-0.0001E_hS_v-0.0043E_hE_v+0.0040E_hI_v+0.0003I_hJ_h-0.3293I_hT_h$ \\
        & $\quad\;\;-0.0086J_hR_h+0.0003J_hS_v+0.0043J_hE_v-0.0001J_hI_v$ \\
        & $\dot{I}_h=-0.0002I_h^2-0.0004V_hI_h+0.0005V_hT_h-0.0033E_hI_h-0.0014E_hJ_h-0.0006I_hT_h$ \\
        & $\dot{J}_h=-0.0010S_h-0.0009R_h-0.0008E_v+0.0003V_h^2+0.0001E_h^2-0.0002I_h^2-0.0001J_h^2$ \\
        & $\quad\;\;+0.0012T_h^2+0.0782S_hE_h-0.0741S_hJ_h-0.0005V_hE_h+0.0082V_hI_h+0.0011V_hJ_h$ \\
        & $\quad\;\;-0.0015V_hT_h+0.0413E_hI_h-0.0061E_hJ_h+0.0001E_hR_h-0.0004I_hJ_h-0.0356I_hT_h$ \\
        & $\quad\;\;+0.0003J_hR_h$ \\
        & $\dot{T}_h=-0.0001T_h^2+0.0002S_hE_h-0.0002S_hJ_h+0.0037V_hI_h-0.0001V_hT_h-0.0086E_hI_h$ \\
        & $\quad\;\;-0.0028I_hT_h$ \\
        \textbf{10\% noise} & $\dot{R}_h=0.0127S_h-0.0005R_h+0.0024E_v-0.0047V_h^2-0.0001I_h^2-0.0001J_h^2-0.0029T_h^2$ \\
        & $\quad\;\;+0.0122S_hE_h+0.1170S_hJ_h-0.0003S_hR_h-0.0009S_hE_v-0.0001V_hE_h+0.0043V_hI_h$ \\
        & $\quad\;\;-0.0002V_hJ_h-0.0369V_hT_h-0.1153E_hI_h+0.0229E_hJ_h-0.0001E_hR_h-0.0003E_hE_v$ \\
        & $\quad\;\;+0.0638I_hT_h+0.0007J_hR_h-0.0011J_hE_v$ \\
        & $\dot{S}_v=19.4872S_h-0.2197R_h-0.0004S_v+0.2936E_v+0.0724I_v-0.0068S_h^2+1.1518V_h^2$ \\
        & $\quad\;\;+0.0372E_h^2-0.4306I_h^2-0.0098J_h^2-0.2813T_h^2-0.0005R_h^2-0.0001S_v^2-0.0004E_v^2$ \\
        & $\quad\;\;-0.0001I_v^2-14.8475S_hE_h+2.9486S_hJ_h-0.0706S_hR_h+0.0006S_hS_v+0.1045S_hE_v$ \\
        & $\quad\;\;-0.0032S_hI_v+0.1591V_hE_h-4.8922V_hI_h-0.1427V_hJ_h-7.3636V_hT_h+0.0170V_hR_h$ \\
        & $\quad\;\;-0.0003V_hS_v-0.0009V_hE_v+0.0002V_hI_v+17.5942E_hI_h+2.5529E_hJ_h-0.0640E_hR_h+$ \\
        & $\quad\;\;0.0031E_hS_v+0.0173E_hE_v-0.0004E_hI_v+0.1353I_hJ_h-3.9490I_hT_h+0.0070I_hR_h$ \\
        & $\quad\;\;-0.0003I_hS_v-0.0008I_hE_v-0.0005I_hI_v+0.2966J_hR_h-0.0033J_hS_v+0.0647J_hE_v$ \\
        & $\quad\;\;+0.0172J_hI_v-0.0001T_hE_v$ \\
        & $\dot{E}_v=0.5364S_h+0.0620R_h-0.0030S_v-0.1168E_v+0.0017I_v+0.2424V_h^2-0.0169E_h^2$ \\
        & $\quad\;\;+0.0769I_h^2+0.0179J_h^2-0.2395T_h^2-0.0004R_h^2-0.0003E_v^2+2.0707S_hE_h-5.7732S_hJ_h$ \\
        & $\quad\;\;+0.0181S_hR_h-0.0017S_hS_v+0.1116S_hE_v-0.0015S_hI_v+0.0142V_hE_h+1.4644V_hI_h$ \\
        & $\quad\;\;-0.0012V_hJ_h+1.6498V_hT_h-0.0046V_hR_h+0.0001V_hS_v-0.0011V_hE_v-0.0001V_hI_v$ \\
        & $\quad\;\;-11.8825E_hI_h-2.3655E_hJ_h+0.0240E_hR_h-0.0005E_hS_v+0.0139E_hE_v+0.0004E_hI_v$ \\
        & $\quad\;\;+0.0480I_hJ_h-0.1349I_hT_h-0.0041I_hR_h+0.0003I_hS_v-0.0005I_hE_v+0.0001I_hI_v$ \\
        & $\quad\;\;+0.0126J_hR_h+0.0093J_hE_v-0.0066J_hI_v+0.0001T_hR_h-0.0001T_hE_v$ \\
        & $\dot{I}_v=0.3300E_v-0.0714I_v$ \\
        \hline\hline
        & 
        $\dot{S}_h=0.0771S_h+1.0085E_h-1.6372J_h+0.0036R_h+0.0002E_v+0.0003I_v+0.0066V_h^2$ \\
        & $\quad\;\;-0.0007E_h^2-0.0411I_h^2+0.0011J_h^2+0.0089T_h^2+0.0649S_hE_h-9.3178S_hI_h-0.1007S_hJ_h$ \\
        & $\quad\;\;+0.0013S_hR_h-0.0001S_hS_v+0.0004S_hE_v+0.0002S_hI_v+0.0026V_hE_h+0.1928V_hI_h$ \\
        \textbf{20\% noise} & $\quad\;\;-0.0069V_hJ_h-0.2055V_hT_h+2.1215E_hI_h+0.1225E_hJ_h+2.7352E_hT_h-0.0102E_hR_h$ \\
        & $\quad\;\;+0.0006E_hS_v+0.0024E_hE_v-0.0033E_hI_v+0.0052I_hJ_h-0.3482I_hT_h+0.0191J_hR_h$ \\
        & $\quad\;\;-0.0005J_hS_v+0.0009J_hE_v+0.0023J_hI_v$ \\
    \end{tabularx}
\end{table}
\begin{table}[ht]
    \begin{tabularx}{1\linewidth}{|l|X|} 
        & $\dot{V}_h=0.0001S_h+0.0003E_h-0.0005J_h-0.0001S_hE_h-0.0020S_hI_h-0.0003V_hT_h$ \\
        & $\quad\;\;-0.0002E_hI_h+0.0001E_hJ_h+0.0003E_hT_h+0.0001I_hT_h$ \\
        & $\dot{E}_h=-0.0690S_h-1.2195E_h+1.7333J_h-0.0028R_h-0.0001E_v-0.0002I_v-0.0051V_h^2$ \\
        & $\quad\;\;+0.0008E_h^2+0.0415I_h^2-0.0011J_h^2-0.0062T_h^2-0.0773S_hE_h+9.1702S_hI_h+0.1148S_hJ_h$ \\
        & $\quad\;\;-0.0013S_hR_h+0.0001S_hS_v-0.0004S_hE_v-0.0002S_hI_v-0.0024V_hE_h-0.1948V_hI_h$ \\
        & $\quad\;\;+0.0065V_hJ_h+0.2153V_hT_h-2.0853E_hI_h-0.1223E_hJ_h-2.8764E_hT_h+0.0103E_hR_h$ \\
        & $\quad\;\;-0.0006E_hS_v-0.0024E_hE_v+0.0033E_hI_v-0.0050I_hJ_h+0.3157I_hT_h-0.0185J_hR_h$ \\
        & $\quad\;\;+0.0005J_hS_v-0.0008J_hE_v-0.0023J_hI_v$ \\
        & $\dot{I}_h=-0.0001I_h^2-0.0009S_hI_h-0.0001V_hI_h+0.0002V_hT_h-0.0154E_hI_h-0.0003E_hJ_h$ \\
        & $\quad\;\;-0.0219E_hT_h-0.0004I_hT_h$ \\
        & $\dot{J}_h=0.2000E_h-0.2010J_h$ \\
        & $\dot{T}_h=0.0001I_h^2-0.0002T_h^2+0.0187S_hI_h+0.0002V_hT_h+0.0026E_hI_h+0.0003E_hJ_h$ \\
        & $\quad\;\;-0.0012E_hT_h-0.0007I_hT_h$ \\
        \textbf{20\% noise} & $\dot{R}_h=0.2000J_h+0.0001I_h^2+0.0002T_h^2+0.0066S_hI_h+0.0003V_hI_h-0.0005V_hT_h$ \\
        & $\quad\;\;+0.0061E_hI_h+0.0198E_hT_h+0.0009I_hT_h$ \\
        & $\dot{S}_v=13.9995S_h+8.3165E_h-107.6831J_h+1.5637R_h+0.0021S_v+0.7773E_v+0.3135I_v$ \\
        & $\quad\;\;-0.0035S_h^2+0.3382V_h^2+0.0222E_h^2-0.6528I_h^2-0.0041J_h^2-0.2177T_h^2-0.0001R_h^2$ \\
        & $\quad\;\;-0.0001S_v^2+0.0001E_v^2-0.0002I_v^2-3.5709S_hE_h+13.8755S_hI_h+5.9628S_hJ_h$ \\
        & $\quad\;\;-0.0009S_hS_v-0.0981S_hE_v+0.0093S_hI_v+0.0152V_hE_h-1.5859V_hI_h-0.0961V_hJ_h$ \\
        & $\quad\;\;-9.1878V_hT_h+0.0062V_hR_h-0.0002V_hS_v+0.0014V_hE_v+51.7889E_hI_h+10.8971E_hJ_h$ \\
        & $\quad\;\;+132.0941E_hT_h-0.1361E_hR_h+0.0039E_hS_v-0.0938E_hE_v-0.0134E_hI_v+0.0752I_hJ_h$ \\
        & $\quad\;\;+21.6784I_hT_h+0.0003I_hR_h+0.0005I_hS_v+0.0004I_hE_v+0.0004I_hI_v-0.1554J_hR_h$ \\
        & $\quad\;\;+0.0032J_hS_v-0.1228J_hE_v-0.0161J_hI_v-0.0001T_hR_h+0.0001T_hE_v+0.0129S_hR_h$ \\
        & $\dot{E}_v=0.2835S_h+2.2892E_h-5.4470J_h+0.0603R_h-0.0008S_v-0.0059E_v+0.0007I_v$ \\
        & $\quad\;\;+0.0001S_h^2-0.1145V_h^2-0.0078E_h^2-0.0407I_h^2+0.0130J_h^2-0.2650T_h^2-0.0004E_v^2$ \\
        & $\quad\;\;+0.9185S_hE_h-3.9999S_hI_h-3.0961S_hJ_h-0.0649S_hR_h+0.0017S_hS_v+0.0828S_hE_v$ \\
        & $\quad\;\;-0.0084S_hI_v-0.0293V_hE_h+0.9885V_hI_h+0.1435V_hJ_h-1.4628V_hT_h-0.0044V_hR_h$ \\
        & $\quad\;\;+0.0001V_hS_v-0.0019V_hE_v-1.2244E_hI_h-1.7179E_hJ_h+38.1332E_hT_h+0.0160E_hR_h$ \\
        & $\quad\;\;+0.0001E_hS_v+0.0152E_hE_v-0.0012E_hI_v-0.1147I_hJ_h+3.9685I_hT_h-0.0044I_hR_h$ \\
        & $\quad\;\;+0.0003I_hS_v+0.0005I_hE_v+0.0005I_hI_v-0.1054J_hR_h+0.0035J_hS_v+0.0036J_hE_v$ \\
        & $\quad\;\;-0.0143J_hI_v+0.0002T_hR_h-0.0001T_hE_v$ \\
        & $\dot{I}_v=0.3300E_v-0.0714I_v$ \\
        \hline\hline
        & $\dot{S}_h=-0.0579E_h-0.0208J_h+0.0055R_h+0.0007E_h^2-0.0029I_h^2-0.0013J_h^2+0.0380T_h^2$ \\
        & $\quad\;\;+1.4002S_hV_h-0.1158S_hE_h-14.0361S_hI_h+0.1636S_hJ_h-0.0022S_hR_h+0.2436V_hI_h$ \\
        & $\quad\;\;+0.0007V_hJ_h-0.5163V_hT_h-0.8032E_hI_h-0.6265E_hJ_h+4.7516E_hT_h+0.0157E_hR_h$ \\
        & $\quad\;\;+0.0183E_hE_v-0.0036E_hI_v-0.0007I_hJ_h+1.9407I_hT_h-24.2167J_hT_h-0.0110J_hR_h$ \\
        & $\quad\;\;-0.0232J_hE_v$ \\
        & $\dot{V}_h=0.0441J_hT_h$ \\
        & $\dot{E}_h=0.0680E_h-0.0566J_h+0.0016I_h^2-0.0226T_h^2+0.3164S_hV_h+0.0285S_hE_h$ \\
        & $\quad\;\;+13.8928S_hI_h-0.0526S_hJ_h-0.2788V_hI_h-0.0420V_hT_h+0.7916E_hI_h+0.6544E_hJ_h$ \\
        & $\quad\;\;-3.2952E_hT_h-0.0154E_hR_h-0.0180E_hE_v+0.0036E_hI_v-1.0337I_hT_h+16.9231J_hT_h$ \\
        & $\quad\;\;+0.0043J_hR_h+0.0223J_hE_v$ \\
        & $\dot{I}_h=-0.0883S_hI_h-0.0246E_hI_h-0.0497E_hT_h-0.0338J_hT_h$ \\
        \textbf{25\% noise} & $\dot{J}_h=0.2000E_h-0.2010J_h$ \\
        & $\dot{T}_h=0.0085S_hI_h+0.0090E_hI_h+0.0295E_hT_h-0.0914J_hT_h$ \\
        & $\dot{R}_h=0.0014E_h+0.1984J_h-0.0141S_hV_h+0.0817S_hI_h+0.0148E_hI_h+0.0190E_hT_h+0.1264J_hT_h$ \\
        & $\dot{S}_v=12.0310S_h-48.7453E_h+22.9846J_h+1.5582R_h-0.0276S_v+0.0746E_v+0.0740I_v$ \\
        & $\quad\;\;-0.0083S_h^2+2.8666V_h^2+0.0647E_h^2+0.0222I_h^2-0.1099J_h^2+4.4383T_h^2-0.0036R_h^2$ \\
        & $\quad\;\;+83.3180S_hV_h-30.9488S_hE_h-289.5507S_hI_h+31.3275S_hJ_h-0.0514S_hR_h$ \\
        & $\quad\;\;-0.0100S_hS_v+0.0136S_hE_v-0.0193S_hI_v+0.0569V_hE_h-1.1562V_hI_h-0.0063V_hJ_h$ \\
        & $\quad\;\;-6.2468V_hT_h+0.0295V_hR_h+0.0038V_hE_v+55.1224E_hI_h+9.6804E_hJ_h$ \\
        & $\quad\;\;-191.1162E_hT_h+0.0893E_hR_h+0.0011E_hS_v-0.0324E_hE_v-0.0128E_hI_v$ \\
        & $\quad\;\;+0.0228I_hJ_h+10.8828I_hT_h-0.0207I_hR_h-0.0045I_hE_v-264.6920J_hT_h+1.2324J_hR_h$ \\
        & $\quad\;\;-0.0385J_hS_v-0.0255J_hE_v+0.0059J_hI_v$ \\
    \end{tabularx}
\end{table}
\begin{table}[ht]
    \begin{tabularx}{1\linewidth}{|l|X|} 
        & $\dot{E}_v=-0.8714S_h+15.1611E_h+40.6121J_h-0.2658R_h+0.0051S_v-0.4037E_v$ \\
        & $\quad\;\;+0.2627V_h^2-0.0243E_h^2+0.1464I_h^2+0.0143J_h^2-0.4236T_h^2-210.1857S_hV_h$ \\
        & $\quad\;\;+8.8759S_hE_h-88.3987S_hI_h-1.6221S_hJ_h+0.2491S_hR_h-0.0237S_hE_v-0.0068S_hI_v$ \\
        \textbf{25\% noise} & $\quad\;\;-0.0586V_hE_h+2.2050V_hI_h+0.0199V_hJ_h+3.6401V_hT_h-0.0089V_hR_h-12.1838E_hI_h$ \\
        & $\quad\;\;-3.7597E_hJ_h+27.8119E_hT_h+0.0668E_hR_h+0.0116E_hE_v-0.0659I_hJ_h-9.9347I_hT_h$ \\
        & $\quad\;\;-0.0069I_hR_h+86.7477J_hT_h-0.1953J_hR_h+0.0061J_hS_v+0.0200J_hE_v+0.0104J_hI_v$ \\
        & $\dot{I}_v=0.3300E_v-0.0714I_v$ \\
        \hline\hline
        & $\dot{S}_h=-0.0156S_h+0.0714E_h+0.0285J_h+0.0031R_h-0.0008E_v-0.0042V_h^2-0.0094I_h^2$ \\
        & $\quad\;\;+0.0330T_h^2+0.9301S_hV_h-0.0722S_hE_h-4.0594S_hI_h-0.0124S_hJ_h+0.1990V_hI_h$ \\
        & $\quad\;\;-0.1401V_hT_h-0.4292E_hI_h+0.3518E_hJ_h+0.7964E_hT_h+0.0040E_hR_h-0.0022E_hI_v$ \\
        & $\quad\;\;-0.2542I_hT_h-10.0913J_hT_h-0.0016J_hR_h+0.0038J_hE_v$ \\
        & $\dot{V}_h=0.0442J_hT_h$ \\
        & $\dot{E}_h=-0.1892E_h-0.0226J_h+0.0032R_h+0.0008E_v+0.0117V_h^2+0.0112I_h^2-0.0187T_h^2$ \\
        & $\quad\;\;+0.0396S_hE_h+3.4686S_hI_h+0.0081S_hJ_h-0.0024S_hR_h-0.2020V_hI_h$ \\
        & $\quad\;\;+0.3935S_hV_h+0.1138V_hT_h+0.4713E_hI_h-0.3503E_hJ_h-1.2770E_hT_h-0.0040E_hR_h$ \\
        & $\quad\;\;+0.0023E_hI_v+0.2733I_hT_h+7.3762J_hT_h-0.0038J_hE_v$ \\
        & $\dot{I}_h=-0.0896S_hI_h-0.0245E_hI_h-0.0500E_hT_h-0.0342J_hT_h$ \\
        & $\dot{J}_h=0.2000E_h-0.2010J_h$ \\
        & $\dot{T}_h=0.0220S_hI_h+0.0090E_hI_h+0.0285E_hT_h-0.0964J_hT_h$ \\
        & $\dot{R}_h=0.1992J_h+0.0912S_hI_h+0.0155E_hI_h+0.0160E_hT_h+0.1356J_hT_h$ \\
        \textbf{30\% noise} & $\dot{S}_v=11.7830S_h-51.8742E_h+24.3361J_h+1.5721R_h-0.0280S_v+0.1057E_v+0.0716I_v$ \\
        & $\quad\;\;-0.0074S_h^2+2.2438V_h^2+0.0653E_h^2-0.0115I_h^2-0.1117J_h^2+3.7532T_h^2-0.0032R_h^2$ \\
        & $\quad\;\;+179.9953S_hV_h-27.6502S_hE_h-225.2304S_hI_h+30.1578S_hJ_h-0.1533S_hR_h$ \\
        & $\quad\;\;-0.0097S_hS_v-0.0175S_hI_v-3.6124V_hI_h+0.0952V_hJ_h-0.1879V_hT_h+0.0288V_hR_h$ \\
        & $\quad\;\;+0.0039V_hE_v+68.0251E_hI_h+7.3881E_hJ_h-172.2410E_hT_h+0.0654E_hR_h$ \\
        & $\quad\;\;+0.0018E_hS_v-0.0121E_hE_v-0.0138E_hI_v-0.0510I_hJ_h+14.7902I_hT_h-0.0194I_hR_h$ \\
        & $\quad\;\;-0.0043I_hE_v-176.6532J_hT_h+1.0070J_hR_h-0.0351J_hS_v-0.0781J_hE_v-0.0061J_hI_v$ \\
        & $\dot{E}_v=-0.9268S_h+19.3397E_h+38.5117J_h-0.2671R_h+0.0050S_v-0.4005E_v-0.0007I_v$ \\
        & $\quad\;\;+0.2876V_h^2-0.0292E_h^2+0.3125I_h^2+0.0199J_h^2-0.3599T_h^2-243.0764S_hV_h$ \\
        & $\quad\;\;+10.2161S_hE_h-84.1048S_hI_h-3.3197S_hJ_h+0.2887S_hR_h-0.0278S_hE_v-0.0066S_hI_v$ \\
        & $\quad\;\;-0.0572V_hE_h+2.7380V_hI_h-0.0029V_hJ_h+3.4238V_hT_h-0.0106V_hR_h-31.1892E_hI_h$ \\
        & $\quad\;\;-3.6842E_hJ_h-13.1911E_hT_h+0.0498E_hR_h+0.0098E_hE_v+0.0016E_hI_v-0.0440I_hJ_h$ \\
        & $\quad\;\;-11.4701I_hT_h-0.0061I_hR_h+16.1899J_hT_h-0.1492J_hR_h+0.0060J_hS_v+0.0300J_hE_v$ \\
        & $\quad\;\;+0.0086J_hI_v$ \\
        & $\dot{I}_v=0.3300E_v-0.0714I_v$ \\
        \hline\hline
        & $\dot{S}_h=-0.0574S_h+0.0166E_h+100.1913I_h-0.0559J_h+0.0088R_h+0.0169V_h^2-0.2219I_h^2$ \\
        & $\quad\;\;+0.1462T_h^2+1.1089S_hV_h-0.0451S_hE_h+16.2672S_hI_h+0.0051S_hJ_h-41.4213S_hT_h$ \\
        & $\quad\;\;-0.0029S_hR_h+0.4514V_hI_h-0.4274V_hT_h+12.1288E_hI_h-0.3814E_hJ_h$ \\
        & $\quad\;\;+16.7916E_hT_h-0.0076E_hR_h+0.0064E_hE_v-0.0047E_hI_v+0.8060I_hT_h$ \\
        & $\quad\;\;-12.9281J_hT_h+0.0154J_hR_h-0.0091J_hE_v$ \\
        & $\dot{V}_h=0.0113I_h-0.0031S_hI_h+0.0565S_hT_h+0.0090J_hT_h$ \\
        & $\dot{E}_h=0.0570S_h-0.1896E_h-97.8601I_h+0.0953J_h-0.0028R_h-0.0175V_h^2+0.2203I_h^2$ \\
        & $\quad\;\;-0.1363T_h^2+1.3290S_hV_h+0.0298S_hE_h-15.7228S_hI_h-0.0219S_hJ_h+39.7789S_hT_h$ \\
        & $\quad\;\;-0.4548V_hI_h+0.4301V_hT_h-12.1019E_hI_h+0.3722E_hJ_h-17.2771E_hT_h+0.0091J_hE_v$ \\
        & $\quad\;\;+0.0068E_hR_h-0.0064E_hE_v+0.0047E_hI_v-0.8038I_hT_h+10.3792J_hT_h-0.0154J_hR_h$ \\
        & $\dot{I}_h=-0.2439I_h$ \\
        & $\dot{J}_h=0.2000E_h-0.2010J_h$ \\
        \textbf{40\% noise} & $\dot{T}_h=0.0977I_h-0.0772S_hT_h-0.0101E_hT_h-0.0554J_hT_h$ \\
        & $\dot{R}_h=0.1300I_h+0.1991J_h+0.1490S_hT_h+0.0011E_hI_h+0.0252E_hT_h+0.0206J_hT_h$ \\
        & $\dot{S}_v=4.5246S_h+14.6332E_h+2017.7062I_h-24.5532J_h+1.4608R_h-0.0263S_v$ \\
        & $\quad\;\;+0.0715E_v+0.0629I_v-1.0930V_h^2-0.0112E_h^2-0.8709I_h^2+2.5308T_h^2$ \\
        & $\quad\;\;+519.1415S_hV_h-2.4979S_hE_h+239.8974S_hI_h+2.5444S_hJ_h-118.9400S_hT_h$ \\
        & $\quad\;\;-0.4937S_hR_h-0.0056S_hS_v+0.0100S_hE_v-0.0094S_hI_v-0.0415V_hE_h-1.7474V_hI_h$ \\
        & $\quad\;\;+0.0047V_hJ_h+6.2585V_hT_h+0.0022V_hR_h-51.5012E_hI_h+0.7245E_hJ_h$ \\
        & $\quad\;\;-158.8168E_hT_h-0.2023E_hR_h-0.0118E_hS_v-0.0450E_hE_v-0.0080E_hI_v-0.0299J_hI_v$ \\
        & $\quad\;\;-0.0170I_hJ_h-2.2124I_hT_h+0.0027I_hR_h-523.2695J_hT_h-0.0908J_hR_h+0.0436J_hE_v$ \\
    \end{tabularx}
\end{table}
\begin{table}[ht]
    \begin{tabularx}{1\linewidth}{|l|X|} 
        & $\dot{E}_v=-0.8558S_h+14.6920E_h+203.9968I_h+40.6010J_h-0.2953R_h+0.0057S_v$ \\
        & $\quad\;\;-0.4006E_v+0.2266V_h^2-0.0209E_h^2+0.1245I_h^2+0.0113J_h^2+0.0110T_h^2-239.3402S_hV_h$ \\
        & $\quad\;\;+7.4810S_hE_h-51.3426S_hI_h-0.1662S_hJ_h+94.5308S_hT_h+0.2846S_hR_h-0.0272S_hE_v$ \\
        \textbf{40\% noise} & $\quad\;\;-0.0068S_hI_v-0.0174V_hE_h+1.6476V_hI_h-0.0510V_hJ_h+2.5020V_hT_h-0.0084V_hR_h$ \\
        & $\quad\;\;-31.0530E_hI_h-4.3047E_hJ_h-35.9685E_hT_h+0.0100E_hE_v-0.0300I_hJ_h-7.1400I_hT_h$ \\
        & $\quad\;\;-0.0076I_hR_h-74.8799J_hT_h-0.0412J_hR_h+0.0325J_hE_v$ \\
        & $\dot{I}_v=0.3300E_v-0.0714I_v$ \\
        \hline\hline
        & $\dot{S}_h=-0.0681S_h+0.5799V_h-0.0617E_h+85.7939I_h+0.2986J_h+0.0019R_h+0.0001S_v$ \\
        & $\quad\;\;+0.0010E_v-0.0006I_v+0.0145V_h^2-0.0629I_h^2-0.0006J_h^2+0.0283T_h^2-0.6514S_hV_h$ \\
        & $\quad\;\;-0.0414S_hE_h+11.5635S_hI_h+0.0682S_hJ_h-25.7718S_hT_h-0.0006S_hR_h$ \\
        & $\quad\;\;-0.0005S_hE_v+0.0005V_hE_h+0.2284V_hI_h-0.0003V_hJ_h-0.0978V_hT_h-2.6595E_hI_h$ \\
        & $\quad\;\;+0.0139E_hJ_h-1.5883E_hT_h-0.0078E_hR_h-0.0005E_hS_v-0.0021E_hE_v-0.0012E_hI_v$ \\
        & $\quad\;\;-0.0008I_hJ_h+0.3520I_hT_h-0.0001I_hR_h-8.7231J_hT_h+0.0136J_hR_h+0.0001J_hS_v$ \\
        & $\quad\;\;+0.0070J_hE_v-0.0032J_hI_v$ \\
        & $\dot{V}_h=0.0001S_h-0.0067V_h-0.0001E_h-0.0016I_h-0.0002J_h+0.0007S_hI_h-0.0027S_hT_h$ \\
        & $\quad\;\;-0.0001V_hT_h+0.0001E_hT_h+0.0001I_hT_h-0.0045J_hT_h$ \\
        & $\dot{E}_h=0.0653S_h+2.7041V_h-0.1379E_h-83.7606I_h-0.2990J_h-0.0030R_h-0.0010E_v$ \\
        & $\quad\;\;+0.0006I_v-0.0141V_h^2+0.0628I_h^2+0.0006J_h^2-0.0261T_h^2+0.4954S_hV_h+0.0440S_hE_h$ \\
        & $\quad\;\;-11.7921S_hI_h-0.0748S_hJ_h+25.1371S_hT_h+0.0010S_hR_h+0.0005S_hE_v+0.0032J_hI_v$ \\
        & $\quad\;\;-0.0005V_hE_h-0.2280V_hI_h+0.0002V_hJ_h+0.1004V_hT_h+2.5722E_hI_h-0.0151E_hJ_h$ \\
        & $\quad\;\;+1.4119E_hT_h+0.0079E_hR_h+0.0005E_hS_v+0.0021E_hE_v+0.0011E_hI_v+0.0008I_hJ_h$ \\
        & $\quad\;\;-0.3501I_hT_h+0.0001I_hR_h+8.7796J_hT_h-0.0132J_hR_h-0.0001J_hS_v-0.0070J_hE_v$ \\
        & $\dot{I}_h=-0.2439I_h$ \\
        & $\dot{J}_h=0.2000E_h-0.2010J_h$ \\
        & $\dot{T}_h=0.0002V_h+0.0852I_h-0.0001S_hV_h+0.0075S_hI_h-0.0734S_hT_h+0.0012E_hI_h$ \\
        & $\quad\;\;-0.0070E_hT_h-0.0586J_hT_h$ \\
        \textbf{50\% noise} & $\dot{R}_h=0.0018V_h+0.1575I_h+0.2000J_h-0.0002S_hV_h-0.0074S_hI_h+0.0716S_hT_h$ \\
        & $\quad\;\;-0.0012E_hI_h+0.0070E_hT_h+0.0584J_hT_h$ \\
        & $\dot{S}_v=11.5742S_h-98.6007V_h-10.0079E_h+504.9588I_h+39.6536J_h+1.8372R_h$ \\
        & $\quad\;\;-0.0109S_v+0.0441E_v+0.0522I_v-0.0012S_h^2+0.5215V_h^2+0.0154E_h^2-0.6016I_h^2$ \\
        & $\quad\;\;-0.0048J_h^2-1.6275T_h^2-0.0008R_h^2-0.0001S_v^2-0.0001E_v^2-0.0001I_v^2-176.3566S_hV_h$ \\
        & $\quad\;\;+1.5625S_hE_h-13.0573S_hI_h+1.8544S_hJ_h+414.0923S_hT_h+0.5207S_hR_h-0.0077S_hS_v$ \\
        & $\quad\;\;+0.0169S_hE_v-0.0043S_hI_v-0.0030V_hE_h+0.8613V_hI_h-0.0164V_hJ_h-3.3153V_hT_h$ \\
        & $\quad\;\;+0.0006V_hR_h+0.0003V_hE_v+0.0002V_hI_v-0.1135E_hI_h-2.1316E_hJ_h+173.1062E_hT_h$ \\
        & $\quad\;\;-0.0513E_hR_h-0.0007E_hS_v+0.0004E_hE_v-0.0031E_hI_v-0.0085I_hJ_h-13.2927I_hT_h$ \\
        & $\quad\;\;+0.0083I_hR_h-0.0006I_hS_v-0.0004I_hE_v-0.0004I_hI_v+279.3249J_hT_h-0.4763J_hR_h$ \\
        & $\quad\;\;+0.0116J_hS_v+0.0832J_hE_v+0.0159J_hI_v-0.0005T_hR_h$ \\
        & $\dot{E}_v=0.3728S_h+1094.1587V_h-9.3292E_h-206.7940I_h+25.0057J_h-1.0060R_h$ \\
        & $\quad\;\;+0.0018S_v-0.3547E_v+0.0084I_v-0.0010S_h^2-0.0597V_h^2-0.0130E_h^2+0.2387I_h^2$ \\
        & $\quad\;\;-0.0069J_h^2+0.6226T_h^2+0.0002R_h^2+86.5970S_hV_h-0.3654S_hE_h-76.0668S_hI_h$ \\
        & $\quad\;\;+0.4254S_hJ_h-467.7718S_hT_h-0.2824S_hR_h-0.0040S_hS_v-0.0320S_hE_v-0.0062S_hI_v$ \\
        & $\quad\;\;-0.0138V_hE_h+2.1107V_hI_h+0.0169V_hJ_h+5.5149V_hT_h-0.0032V_hR_h+0.0002V_hS_v$ \\
        & $\quad\;\;+4.1111E_hI_h-1.3581E_hJ_h-63.7935E_hT_h+0.0735E_hR_h+0.0004E_hS_v+0.0063E_hE_v$ \\
        & $\quad\;\;+0.0005E_hI_v-0.0307I_hJ_h+0.7028I_hT_h-0.0111I_hR_h+0.0002I_hS_v-60.6075J_hT_h$ \\
        & $\quad\;\;+0.2632J_hR_h-0.0019J_hS_v-0.0213J_hE_v-0.0042J_hI_v+0.0003T_hR_h$ \\
        & $\dot{I}_v=0.3300E_v-0.0714I_v$ \\
        \hline
    \end{tabularx}
\end{table}

\clearpage
\bibliographystyle{plain}
\bibliography{REFS}

@article{wilson,
  title={The law of mass action in epidemiology},
  author={Wilson, Edwin B and Worcester, Jane},
  journal={Proceedings of the National Academy of Sciences},
  volume={31},
  number={1},
  pages={24--34},
  year={1945}
}

@article{brunton,
  title={Discovering governing equations from data by sparse identification of nonlinear dynamical systems},
  author={Brunton, Steven L and Proctor, Joshua L and Kutz, J Nathan},
  journal={Proceedings of the National Academy of Sciences},
  volume={113},
  number={15},
  pages={3932--3937},
  year={2016},
  publisher={National Academy of Sciences}
}

@article{zhang,
  title={On the convergence of the SINDy algorithm},
  author={Zhang, Linan and Schaeffer, Hayden},
  journal={Multiscale Modeling \& Simulation},
  volume={17},
  number={3},
  pages={948--972},
  year={2019},
  publisher={SIAM}
}

@article{morrison,
  title={Reemergence of chikungunya virus},
  author={Morrison, Thomas E},
  journal={Journal of virology},
  volume={88},
  number={20},
  pages={11644--11647},
  year={2014},
  publisher={American Society for Microbiology 1752 N St., NW, Washington, DC}
}

@article{zhang1,
  title={Global resurgence of Chikungunya virus: outbreak drivers and emerging solutions},
  author={Zhang, Yi and Wu, Jing and Cheng, Xiaoyang and Yang, Yuxuan and Wang, Xinyu and Zhao, Xiaoyu and Wang, Xiaoyan and Ouyang, Huiling and Ai, Jingwen and Zhang, Wenhong},
  journal={Emerging Microbes \& Infections},
  volume={15},
  number={1},
  pages={2603714},
  year={2026},
  publisher={Taylor \& Francis}
}

@article{zhao,
  title={Chikungunya Virus in 2025: Epidemiology, Immunopathogenesis, and Vaccine Development—A Narrative Review},
  author={Zhao, Chenxi and Ge, Ziruo and Zhang, Tingyu and Jiang, Zhouling and Tian, Di and Chen, Zhihai},
  journal={Infection and Drug Resistance},
  pages={1--14},
  year={2026},
  publisher={Taylor \& Francis}
}

@article{diekmann,
  title={On the definition and the computation of the basic reproduction ratio R 0 in models for infectious diseases in heterogeneous populations},
  author={Diekmann, Odo and Heesterbeek, Johan Andre Peter and Metz, Johan Anton Jacob},
  journal={Journal of Mathematical Biology},
  volume={28},
  pages={365--382},
  year={1990},
  publisher={Springer}
}

@article{van,
  title={Reproduction numbers and sub-threshold endemic equilibria for compartmental models of disease transmission},
  author={Van den Driessche, Pauline and Watmough, James},
  journal={Mathematical biosciences},
  volume={180},
  number={1-2},
  pages={29--48},
  year={2002},
  publisher={Elsevier}
}

@article{yang,
  title={Basic reproduction numbers for a class of reaction-diffusion epidemic models},
  author={Yang, Chayu and Wang, Jin},
  journal={Bulletin of mathematical biology},
  volume={82},
  number={8},
  pages={111},
  year={2020},
  publisher={Springer}
}

@article{lenfesty,
  title={Uncovering dynamical equations of stochastic decision models using data-driven SINDy algorithm},
  author={Lenfesty, Brendan and Bhattacharyya, Saugat and Wong-Lin, KongFatt},
  journal={Neural Computation},
  volume={37},
  number={3},
  pages={569--587},
  year={2025},
  publisher={MIT Press 255 Main Street, 9th Floor, Cambridge, Massachusetts 02142, USA~…}
}

@article{breda,
  title={Sparse identification of nonlinear dynamics for stochastic delay differential equations},
  author={Breda, Dimitri and Conte, Dajana and D’Ambrosio, Raffaele and Santaniello, Ida and Tanveer, Muhammad},
  journal={Journal of Computational and Applied Mathematics},
  pages={117247},
  year={2025},
  publisher={Elsevier}
}

@article{fukami,
  title={Sparse identification of nonlinear dynamics with low-dimensionalized flow representations},
  author={Fukami, Kai and Murata, Takaaki and Zhang, Kai and Fukagata, Koji},
  journal={Journal of Fluid Mechanics},
  volume={926},
  pages={A10},
  year={2021},
  publisher={Cambridge University Press}
}

@incollection{alrazen,
  title={Machine learning-based new sparse algorithms and reinforcement learning for computational fluid dynamics},
  author={Alrazen, Hayder and Yaacob, Izzat Najmi and Mazlan, Norkhairunnisa and Yidris, Noorfaizal},
  booktitle={Artificial Intelligence for Computational Fluid Dynamics},
  pages={313--343},
  year={2026},
  publisher={Elsevier}
}

@article{selvitella,
  title={On the Effectiveness of Sparse Identification Methods to Detect Nonlinear Models of Oscillatory Dynamics in Psychology and the Life Sciences.},
  author={Selvitella, Alessandro Maria and Allen, Elliot},
  journal={Nonlinear Dynamics, Psychology \& Life Sciences},
  volume={30},
  number={1},
  year={2026}
}

@article{muhammed,
  title={Data-driven identification of biological systems using multi-scale analysis},
  author={Muhammed, Ismaila and Manias, Dimitris M and Goussis, Dimitris A and Hatzikirou, Haralampos},
  journal={PLOS Computational Biology},
  volume={21},
  number={11},
  pages={e1013193},
  year={2025},
  publisher={Public Library of Science San Francisco, CA USA}
}

@article{prabhu,
  title={Derivative-free domain-informed data-driven discovery of sparse kinetic models},
  author={Prabhu, Siddharth and Kosir, Nick and Kothare, Mayuresh V and Rangarajan, Srinivas},
  journal={Industrial \& Engineering Chemistry Research},
  volume={64},
  number={5},
  pages={2601--2615},
  year={2025},
  publisher={ACS Publications}
}

@article{guo,
  title={Uncertainty quantification in reduced-order gas-phase atmospheric chemistry modeling using ensemble SINDy},
  author={Guo, Lin and Yang, Xiaokai and Zheng, Zhonghua and Riemer, Nicole and Tessum, Christopher W},
  journal={Journal of Geophysical Research: Machine Learning and Computation},
  volume={1},
  number={4},
  pages={e2024JH000358},
  year={2024},
  publisher={Wiley Online Library}
}

@article{jiang,
  title={Modeling and prediction of the transmission dynamics of COVID-19 based on the SINDy-LM method},
  author={Jiang, Yu-Xin and Xiong, Xiong and Zhang, Shuo and Wang, Jia-Xiang and Li, Jia-Chun and Du, Lin},
  journal={Nonlinear Dynamics},
  volume={105},
  number={3},
  pages={2775--2794},
  year={2021},
  publisher={Springer}
}

@article{babazadeh,
  title={Automated Model Discovery Based on COVID-19 Epidemiologic Data},
  author={Babazadeh Shareh, Morteza and Kleiner, Florian and B{\"o}hme, Michael and H{\"a}gele, Corinna and Dickmann, Petra and Heintzmann, Rainer},
  journal={medRxiv},
  pages={2026--02},
  year={2026},
  publisher={Cold Spring Harbor Laboratory Press}
}

@article{kaheman,
  title={Automatic differentiation to simultaneously identify nonlinear dynamics and extract noise probability distributions from data},
  author={Kaheman, Kadierdan and Brunton, Steven L and Nathan Kutz, J},
  journal={Machine Learning: Science and Technology},
  volume={3},
  number={1},
  pages={015031},
  year={2022},
  publisher={IOP Publishing}
}

@article{fasel,
  title={Ensemble-SINDy: Robust sparse model discovery in the low-data, high-noise limit, with active learning and control},
  author={Fasel, Urban and Kutz, J Nathan and Brunton, Bingni W and Brunton, Steven L},
  journal={Proceedings of the Royal Society A: Mathematical, Physical and Engineering Sciences},
  volume={478},
  number={2260},
  year={2022},
  publisher={The Royal Society}
}

@article{fasel2,
  title={Ensemble-SINDy},
  author={Fasel, U and Kutz, JN and Brunton, BW and Brunton, SL},
  journal={Proceedings: Mathematical, Physical and Engineering Sciences},
  volume={478},
  number={2260},
  pages={1--20},
  year={2022},
  publisher={JSTOR}
}

@article{kaheman2,
  title={SINDy-PI: a robust algorithm for parallel implicit sparse identification of nonlinear dynamics},
  author={Kaheman, Kadierdan and Kutz, J Nathan and Brunton, Steven L},
  journal={Proceedings of the Royal Society A: Mathematical, Physical and Engineering Sciences},
  volume={476},
  number={2242},
  year={2020},
  publisher={The Royal Society}
}

@article{evensen,
  title={The ensemble Kalman filter: Theoretical formulation and practical implementation},
  author={Evensen, Geir},
  journal={Ocean dynamics},
  volume={53},
  number={4},
  pages={343--367},
  year={2003},
  publisher={Springer}
}

@article{lal,
  title={An application of the ensemble Kalman filter in epidemiological modeling},
  author={Lal, Rajnesh and Huang, Weidong and Li, Zhenquan},
  journal={Plos one},
  volume={16},
  number={8},
  pages={e0256227},
  year={2021},
  publisher={Public Library of Science San Francisco, CA USA}
}

@article{belgraoui,
  title={Parameter estimation in a stochastic SEIHR model of COVID-19 using the EnKF: A case study based on real-world data},
  author={Belgraoui, Youssef and Belmaati, Aziza and Kabil, Mustapha},
  journal={Chaos, Solitons \& Fractals},
  volume={208},
  pages={118225},
  year={2026},
  publisher={Elsevier}
}

@article{abbas,
  title={Joint estimation of hand-foot-mouth disease model and prediction in korea using the ensemble kalman filter},
  author={Abbas, Wasim and Lee, Sieun and Kim, Sangil},
  journal={PLOS Computational Biology},
  volume={21},
  number={4},
  pages={e1012996},
  year={2025},
  publisher={Public Library of Science San Francisco, CA USA}
}

@article{coffey2014chikungunya,
  title={Chikungunya virus--vector interactions},
  author={Coffey, Lark L and Failloux, Anna-Bella and Weaver, Scott C},
  journal={Viruses},
  volume={6},
  number={11},
  pages={4628--4663},
  year={2014},
  publisher={MDPI}
}

@article{guzzetta2020spatial,
  title={Spatial modes for transmission of chikungunya virus during a large chikungunya outbreak in Italy: a modeling analysis},
  author={Guzzetta, Giorgio and Vairo, Francesco and Mammone, Alessia and Lanini, Simone and Poletti, Piero and Manica, Mattia and Rosa, Roberto and Caputo, Beniamino and Solimini, Angelo and Torre, Alessandra Della and others},
  journal={BMC medicine},
  volume={18},
  number={1},
  pages={226},
  year={2020},
  publisher={Springer}
}

@article{rosafalco2024ekf,
  title={EKF--SINDy: Empowering the extended Kalman filter with sparse identification of nonlinear dynamics},
  author={Rosafalco, Luca and Conti, Paolo and Manzoni, Andrea and Mariani, Stefano and Frangi, Attilio},
  journal={Computer Methods in Applied Mechanics and Engineering},
  volume={431},
  pages={117264},
  year={2024},
  publisher={Elsevier}
}

@article{rosafalco2024online,
  title={Online learning in bifurcating dynamic systems via SINDy and Kalman filtering},
  author={Rosafalco, Luca and Conti, Paolo and Manzoni, Andrea and Mariani, Stefano and Frangi, Attilio},
  journal={arXiv preprint arXiv:2411.04842},
  year={2024}
}

@article{zhang2022ensemble,
  title={Ensemble Kalman method for learning turbulence models from indirect observation data},
  author={Zhang, Xin-Lei and Xiao, Heng and Luo, Xiaodong and He, Guowei},
  journal={Journal of Fluid Mechanics},
  volume={949},
  pages={A26},
  year={2022},
  publisher={Cambridge University Press}
}

\end{document}